\documentclass[pdflatex,sn-mathphys]{sn-jnl}

\jyear{2021}%

 \RequirePackage{tikz}
 \usepackage{tikz}

\theoremstyle{thmstyleone}%
\theoremstyle{thmstyletwo}%

\theoremstyle{thmstylethree}%

\usepackage{graphicx} 
\usepackage{footnote} 
\usepackage{hyperref} 
 \usepackage{multirow} 
 \usepackage{amsmath} 
 \usepackage{amsfonts} 
 \usepackage{amsthm} 
 
 \usepackage{pgfplots}

\pgfplotsset{compat=1.7}

 \usetikzlibrary{matrix}

 \usepgfplotslibrary{groupplots}
\usepackage{algorithm}
\usepackage{algorithmicx}

 \usepackage{subfigure}
 \usepackage{comment}
 \usepackage{color}

\usepackage{threeparttable}
\usepackage{booktabs}

 \usepackage{amssymb}
 \newtheorem{myDef}{Definition}
\usepackage{enumitem}
\setlist[description]{style=nextline}

\begin{document}

\title[ ]{Session-Based Recommendation by Exploiting Substitutable and Complementary Relationships from Multi-behavior Data}


\author[1]{\fnm{Huizi Wu}}\email{wuhuizisufe@gmail.com}

\author[1]{\fnm{Cong Geng}}\email{gcong.leslie@gmail.com}

\author*[1]{\fnm{Hui Fang}}\email{fang.hui@mail.shufe.edu.cn}

\affil[1]{\orgdiv{Research Institute for Interdisciplinary Science and School of Information Management and Engineering}, \orgname{Shanghai University of Finance and Economics}, \orgaddress{ \country{China}}}


\abstract{Session-based recommendation (SR) aims to dynamically recommend items to a user based on a sequence of the most recent user-item interactions. Most existing studies on SR adopt advanced deep learning methods. However, the majority only consider a special behavior type (e.g., click), while those few considering multi-typed behaviors ignore to take full advantage of the relationships between products (items). In this case, the paper proposes a novel approach, called \underline{S}ubstitutable and \underline{C}omplementary \underline{R}elationships from \underline{M}ulti-behavior Data (denoted as SCRM) to better explore the relationships between products for effective recommendation. Specifically, we firstly construct substitutable and complementary graphs based on a user's sequential behaviors in every session by jointly considering `click' and `purchase' behaviors. We then design a denoising network to remove false relationships, and further consider constraints on the two relationships via a particularly designed loss function. Extensive experiments on two e-commerce datasets demonstrate the superiority of our model over state-of-the-art methods, and the effectiveness of every component in SCRM.}




\maketitle

\section{Introduction} \label{sec:introduction}
With the rapid development of e-commerce, session-based recommendation (SR), which aims to predict the next interacted item under a special type of behavior (anonymous session), has played an increasingly critical role.
Compared with the traditional recommendation methods, session-based recommendation considers each session as an ordered sequence. Based on this idea, various machine learning and deep learning techniques have been exploited, including Markov chains  \cite{rendle2010factorizing}, recurrent neural networks (RNN) \cite{hidasi2018recurrent}, attention mechanism-based \cite{zhang2021personal,kang2018self}, and graph neural networks (GNN) \cite{kipf2016semi,velivckovic2017graph}.  

\begin{figure}[htbp]
    \centering
    \includegraphics[width=10cm]{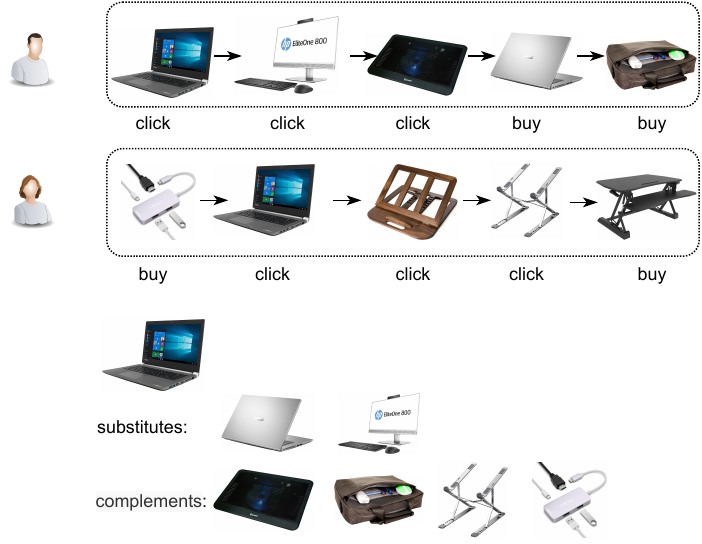}
    \caption{Examples showcasing users' historical interactions.}
    \label{fig:sessionBehaviors}
\end{figure}

Although the aforementioned approaches have achieved encouraging results for session-based recommendation, most of them model a session by treating it as involving only one behavior type (e.g., click or purchase). However, in most scenarios, a session in e-commerce is mixed with different behavior types \cite{fang2020deep}. That is, when a session starts, a user may conduct a series of different actions (such as click and purchase) simultaneously on different products until her needs are finally met. For instance, the upper part of Figure \ref{fig:sessionBehaviors} illustrates two users' historical interactions in a typical e-commerce scenario, where we can observe a clear interconnected relation between items (e.g., substitutes or complements listed in the lower part of Figure \ref{fig:sessionBehaviors}) accompanied with different behavior types (i.e., click and buy) in each sequence. In this case, it is worthwhile to leverage \emph{the sequential dependencies between different behavior types} to better capture the sequential dependencies between different products (items), and thus for more effective next-item prediction.

On the other side, connected products in a session via different behavior types (e.g., co-purchased or co-clicked) might signal different relationships, while substitutable and complementary ones are the most representative. Economic theories \cite{mas1995microeconomic, varian2014intermediate} emphasize that two products are \emph{substitutes} if a user is willing to substitute one product for the other.
In contrast, \emph{complementary} products are inclined to be consumed together, i.e., two products somewhat ``complement" each other.
In other words, \emph{substitutable} products are interchangeable, while \emph{complementary} ones might be purchased together \cite{zhang2018quality,wang2018path}. For instance, when a user wants to buy a t-shirt, she would often retrieve similar t-shirts (i.e., substitutable products) rather than other product types such as food or books. Similarly, after buying a camera, users are more likely to buy films (i.e., complementary products). 
It is thus valuable to explore substitutable and complementary relationships for more effective recommendation.

Previous work might either focus on modeling the features (textual and visual features) of products \cite{mcauley2015inferring,rakesh2019linked}, or utilize graph structure \cite{liu2020decoupled,wang2020beyond} to infer substitutable and complementary relationships between products, but few of them have exerted appropriate constraints between complementary and substitutable relationships towards every same product pair, e.g., a product pair with higher complementary relationship might be less likely to be substitutes. 
Besides, some studies \cite{mcauley2015image,mcauley2015inferring} infer substitutable and complementary relationship by intuitive rules, which might not hold in reality and involve noise. For instance, if users click item $v_i$ also click item $v_j$, the two items $v_i$ and $v_j$ may be substitutable. However, this prior knowledge is not always correct for all sessions. For example in the second session of Figure \ref{fig:sessionBehaviors}, the laptop and notebook stand, although being co-clicked, are complementary rather than substitutable.

To address aforementioned problems, we propose a novel GNN model called SCRM (Session-based Recommendation by Exploiting \underline{S}ubstitutable and \underline{C}omplementary \underline{R}elationships from \underline{M}ulti-behavior Data).
In SCRM, we capture the sequential dependencies between different behavior types by simultaneously considering `click' and `purchase' behaviors in every session\footnote{It should be noted that in this study, we only consider these two behavior types to unveil item relationships since `click' is the most prevalent in e-commerce, whilst `purchase' is the most important and best indicates users' preferences and intents. In the future, more behavior types will be considered.}, with the aim of modeling the substitutable and complementary relationships between items. In particular, we build two graphs to address the substitutable and complementary relationships between products, and adopt denoising network to remove false relationships (noisy ones previously extracted by rules) intelligently. Then, we propose a GNN-based model to learn the final item representation, where on every product pair, appropriate constraints between substitutable and complementary relationships are considered via a particularly designed loss function.  
The contributions are listed as follows: 
\begin{itemize}
\item We are the first to propose a graph neural network model that considers real-time behavior sequences as well as multiple behavior types to infer substitutable and complementary items for session-based recommendation.
\item We design a novel mechanism to learn substitutable and complementary relationships with a careful consideration on removing false relationships of constructed graphs. Besides, on every product pair, the two relationships are appropriately and logically restricted.
\item We conduct extensive experiments on two real-world e-commerce datasets to demonstrate the superiority of SCRM over state-of-the-art approaches, and the effectiveness of every component. Our framework can serve as a guideline for exploiting substitutable and complementary relationships from multi-typed behaviors (not only `click' and `purchase') for more effective session-based recommendation in e-commerce environments.
\end{itemize}
The rest of the paper is organized as follows. Section \ref{sec:literature} summarizes two-fold of related studies: session-based recommendation, and inferring substitutable and complementary relationships. Section \ref{sec:graph} introduces the details of constructing substitutable and complementary graphs from sessions. In Section \ref{sec:model}, we present the architecture of our proposed SCRM, whose effectiveness is thoroughly evaluated in Section \ref{sec:experiments}. Section \ref{sec:conclusions} concludes our study.

\section{Related Work}\label{sec:literature}
Our work is related to two primary areas: session-based recommendation, and the modeling of substitutable and complementary relationships. In the following, we will introduce each part to highlight our contributions over the related studies.

\subsection{Session-based Recommendation}
Session-based recommendation predicts the next item a user will probably like given an anonymous session. Previous studies adopt machine learning techniques that are capable of handling sequential behaviors for session-based recommendation, which is mainly based on Markov chain (MC) \cite{mobasher2002using,wu2013personalized,le2016modeling}. 
For example, Shani et al. \cite{shani2005mdp} propose a novel approach to session-based recommender systems with an Markov Decision Processes (MDP) model. FPMC \cite{rendle2010factorizing} applies matrix factorization and first-order Markov chains to address the sequential dependencies among two adjacent items in a session. Chen et al. \cite{chen2012playlist} treat the music playlist as Markov chain and use a machine learning algorithm, Latent Markov Embedding (LME), to generate the songs' representations. However, MC-based methods are limited to process first-order relationships between items.

Considering the advantages of processing sequential sequences, recurrent neural networks (RNN) \cite{donkers2017sequential, quadrana2017personalizing} and graph neural networks (GNN) \cite{hidasi2016parallel} have been widely adopted in session-based recommendation. In contrast to MC-based methods, RNN-based methods can deal with a much longer sequence. For instance, GRU4Rec \cite{hidasi2015session} is the first method that applies RNN (i.e., a multi-layer gate recurrent unit) to process session data. Later, quite a few RNN-based methods have been proposed.
NARM \cite{li2017neural} explores a hybrid encoder with an attention mechanism to better capture the sequential dependencies among items for more effective session-based recommendation.
STAMP \cite{liu2018stamp} uses multi-layer perceptrons (MLPs) and attention networks to represent long-term and short-term interests within the sequence respectively. In particular, this model can capture users' long-term preferences from the context of the sequence and learn users' current short-term interests from the last clicked product.
Besides, the aforementioned RNN-based methods mainly capture the dependency relationship of items in a session, but ignore to capture item transitions across sessions.

Instead of addressing behavior dependencies in a session as RNN-based methods, GNN-based methods \cite{qiu2019rethinking} can directly capture item relationships across different sessions by learning item representations over session-induced graphs. For instance,
SR-GNN \cite{wu2019session} firstly uses a gated GNN \cite{li2015gated} to encode different sessions into session graphs.
GC-SAN \cite{xu2019graph} models local graph structured dependencies of separated session sequences and designs a multi-layer self-attention network to obtain contextualized non-local representation.
LESSR \cite{chen2020handling} also constructs two graphs (EOP multigraph and shortcut graph) and identify two information loss problems for session-based recommendation, including the lossy session encoding problem and the ineffective long-range dependency capturing problem. DHCN \cite{xia2021self} models sessions as a hyper-graph and then proposes a dual-channel hypergraph convolutional network to improve session-based recommendation. 
GCE-GNN \cite{wang2020global} converts the session sequences into session graphs and constructs a global graph for SR, whilst COTREC \cite{xia2021self1} designs a self-supervised graph co-training framework, which can iteratively select evolving pseudo-labels.

However, besides the inherent problems, all the aforementioned approaches ignore to directly consider the complex and explicit relationships between items, which might lead to insufficient and inaccurate recommendation.

\subsection{Modeling Substitutable and Complementary Relationship}

There are some works which have examined the multi-typed behaviors in e-commerce. For example, in traditional recommendation scenarios, some studies \cite{xia2022multi,xia2021graph,xia2021knowledge} regard the ``purchase” as the target behavior, while other types of behaviors (e.g., click) are referred to as context behaviors. And, they generally design a multi-channel projection mechanism to learn the influence of different behavior types. Besides, in session-based recommendation, MKM-SR \cite{meng2020incorporating} simultaneously incorporates item knowledge (using item attributes) and user micro-behaviors, which contain different operation sequences.
However, the aforementioned studies ignore to explore the complex relationships between products.

On the other hand, previous studies \cite{wan2018representing} have considered exploring the relationships (substitutable and complementary ones) between products in traditional recommendation scenarios (not SR) to improve model accuracy and explainability.
For example, Sceptre \cite{mcauley2015inferring} learns latent topics from textual information to predict substitutable and complementary relationships, whereas Encore \cite{zhang2018quality} combines visual and textual features to mine these relationships, and RSC \cite{zhao2017improving} models them based on co-purchased and co-browsed behaviors. PSMC \cite{wang2018path} learns item embedding representations by identifying useful path constraints to uncover substitutable and complementary relationships.
Triple2vec \cite{wan2018representing} holistically leverages complementarity and compatibility relationships of items, and designs a novel algorithm for product recommendation by adaptively balancing universal product embeddings and users’ product loyalty over time.
A2CF \cite{chen2020try}  extracts attribute information from reviews to model substitutable relationship and then optimizes substitution constraints for recommendation.

As mentioned, quite a set of recent studies have applied GNNs for recommendation, and some of them have also considered to unveil relationships between products. For example, DecGCN \cite{liu2020decoupled} constructs substitutable and complementary graphs via co-viewed and co-purchased products respectively, and learns item representations in separated spaces. However, it infers substitutes and complements separately for traditional recommendation, and ignores to capture the possible dependency between the two types of relationships.

Besides, studies on session-based recommendation like \cite{wang2020beyond} treat the co-clicked products (substitutes) as the side information of the co-purchased products (complements) to facilitate the prediction of purchased items, but they also consider different types of behaviors separately, which might lead to inappropriate modeling and discrimination of the two relationships.
SCG-SPRe \cite{zhang2021learning} also constructs substitutable and complementary graphs as previous studies \cite{mcauley2015inferring} without considering the constraints between different relationships. Besides, it is built for sequential recommendation (where user id is considered, and a sequence involves a user's all historical interactions) instead of session-based recommendation.

Therefore, in our study, we strive to appropriately exploit the dependencies between different behavior types (i.e., click and purchase), and thus address connections between substitutes and complements for effective session-based recommendation.

\section{Converting Sessions to Graphs}\label{sec:graph}
In this section, we firstly formally define our research problem and the major notations. Then, we present how to build substitutable and complementary graphs, as well as the way of measuring the initial edge weights.

\subsection{Problem Definition}
Our task is to predict the next interested item based on a given behavior sequence in chronological order. In particular, let $\mathcal{V}=\{v_1,v_2, \cdots,v_N\}$ denote the item set, where $N$ is the number of items. We further define $\mathcal{B}$ ($b \in \mathcal{B}$) to represent the behavior type, i.e., `click' or `purchase'. 
$\mathcal{S}_f=\{(v_1^S, b_1^S),(v_2^S, b_2^S), \cdots, (v_l^S, b_l^S)\}$ represents the fused session in chronological order, where $v_i^S\in\mathcal{V}$ denotes the $i$-th item being interacted with in session $\mathcal{S}_f$ and $b_i^S$ is the behavior type of $v_i^S$. $l$ denotes the session length. 
Without the loss of generality, we define this research problem as follows:  

\begin{myDef}
Given the fused session $\mathcal{S}_f=\{(v_1^S, b_1^S),(v_2^S, b_2^S), \cdots, (v_l^S, b_l^S)\}$, we generate a probability value for each item $v_i$, $\widehat{y}_i$, and the top $K$ items with the highest values will be recommended.
\end{myDef}
In this paper, matrices are denoted by bold uppercase letters (e.g., $\mathbf{X}$ and $\mathbf{M})$, while the vectors are column vectors and are represented by bold lowercase letters (e.g., $\mathbf{x}$). 

\subsection{Constructing Graphs}
\begin{figure}[htbp]
    \centering
    \includegraphics[width=10cm]{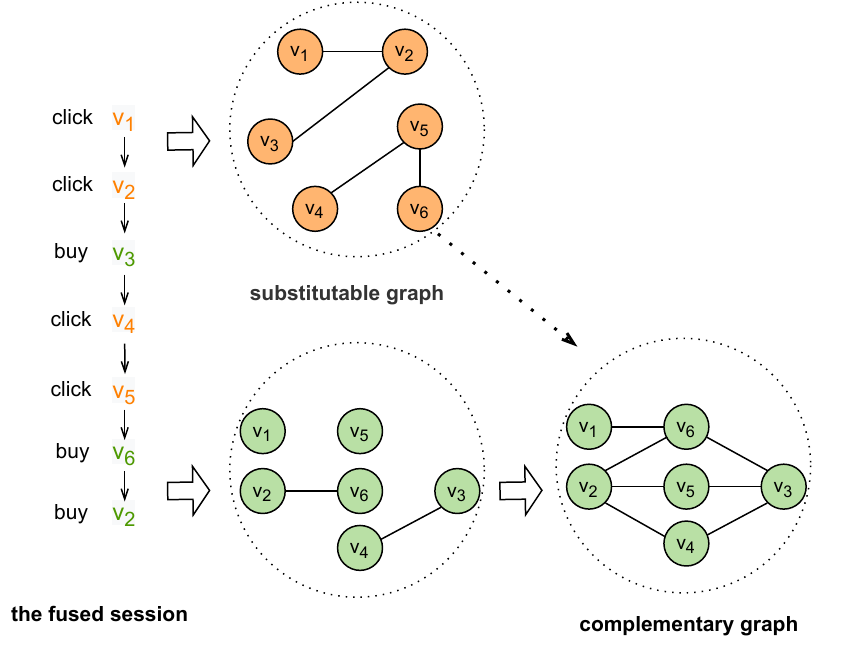}
    \caption{The process of graph construction.}
    \label{fig:Graph1}
\end{figure}

In this study, we aim to explore the dependencies between multiple types of behaviors, i.e., `click' and `purchase', to better exploit the relationships between items, i.e., substitutable and complementary ones, for more effective session-based recommendation.
Therefore, we firstly form the fused session $\mathcal{S}_f$ in chronological order. For example, in Figure \ref{fig:Graph1}, there is a session $\{v_1,v_2,v_3,v_4,v_5,v_6,v_2\}$ and each item in this session is accompanied by one of the two behavior types (i.e., click and purchase). 
We then derive a substitutable graph $\mathcal{G}^s$ and a complementary $\mathcal{G}^c$ from the fused sessions, which are detailed as below. It should be noted that the two graphs are both undirected.

\subsubsection{Forming the substitutable graph.}
Let denote the substitutable graph as $\mathcal{G}^s=(\mathcal{V},\mathcal{E}^s)$, where the edge set $\mathcal{E}^s$ $(E_{i,j}^s \in \mathcal{E}^s)$ summarizes the undirected substitute relationship between items (products). Following \cite{mcauley2015image,mcauley2015inferring}, we firstly label the substitutable product pairs. Particularly, under session-based recommendation scenario, we define two types of substitute relationship between product $v_i$ and $v_j$: (1) $v_i$ 
and $v_j$ are clicked adjacently in session $S_f$; (2) $v_i$ is firstly clicked followed by an immediate purchase of $v_j$. 
In terms of the two principles, in Figure \ref{fig:Graph1}, we construct a substitutable graph where the corresponding product pairs are $v_1 \Leftrightarrow v_2$, $v_2 \Leftrightarrow v_3$, $v_4 \Leftrightarrow v_5$, $v_5 \Leftrightarrow v_6$. Let $w^s_{i,j}$ denote the weight of edge $E^s_{i,j}$, and its initial value is defined by normalizing the corresponding occurred frequency of edge $E^s_{i,j}$ (the frequency of the substitutable relationship, i.e., $v_i\Leftrightarrow v_j$, in all sessions). Note that we only consider the first-order relationship here since: 1) click behavior is quite prevalent in e-commerce; and 2) involving second or higher-order relationships would rather incur noisy connections rather than bring benefits for SR. We will further verify the effectiveness of our choice in Section \ref{sec:experiments}.

\subsubsection{Forming the complementary graph.}
Similarly, let $\mathcal{G}^c=(\mathcal{V},\mathcal{E}^c)$ be
the corresponding complementary graph. We firstly adopt two types of first-order complementary relationship between item $v_i$ and $v_j$ under SR scenario \cite{mcauley2015image}: (1) both $v_i$ and $v_j$ are purchased adjacently in $\mathcal{S}_f$; (2) $v_i$ is firstly purchased succeeded by a click of $v_j$. As shown in Figure \ref{fig:Graph1}, the first-order complement relationships are $v_3 \Leftrightarrow v_4$, $v_6 \Leftrightarrow v_2$.
Since the amount of purchase behavior in e-commerce is rather small compared to that of click behavior, the derived first-order complementary graph is also relatively sparse. In this case, we consider to obtain the second-order complementary relationship to augment the graph. In particular, we consider two intuitively reasonable and robust principles (as demonstrated in Equations \ref{eq:comSecond1} and \ref{eq:comSecond2}) \cite{wang2018path}: 1) if product $v_i$ is complement of $v_k$ and $v_k$ is substitute of $v_j$, then $v_i$ is also a possible complement of $v_j$; 2) if $v_i$ 
and $v_k$ are substitutes while $v_k$ is complement of $v_j$, then we can infer that $v_i$ and $v_j$ is highly probable to have a complementary relationship. 
\begin{equation}\small
(v_i, \text{COM}, v_k) \cap (v_k, \text{SUB}, v_j) \rightarrow (v_i, \text{COM}, v_j)
\label{eq:comSecond1}
\end{equation}
\begin{equation}\small
(v_i, \text{SUB}, v_k) \cap (v_k, \text{COM}, v_j) \rightarrow (v_i, \text{COM}, v_j)
\label{eq:comSecond2}
\end{equation}
Hence, in Figure \ref{fig:Graph1}, we can obtain new edges: $E^c_{2,4}$, $E^c_{3,5}$, $E^c_{2,5}$, $E^c_{1,6}$, and $E^c_{3,6}$, for the complementary graph. Similarly, we define each edge weight $w^c_{i,j}$ and its initial value.

As can be viewed in Equations \ref{eq:comSecond1} and \ref{eq:comSecond2}, 
we can see that the second-order complementary relationships are obtained by using substitutable relationships as bridges. That is, second-order complementary relationships cannot be learned sorely on a complementary graph. However, for the second-order substitutable relationships like ($(v_i, \text{SUB}, v_k) \cap (v_k, \text{SUB}, v_j) \rightarrow (v_i, \text{SUB}, v_j)$), it can be probably learned by a multi-layer graph neural network if needed. This is the third reason why we construct the substitutable graph only by first-order relationship.

\section{The SCRM Model}\label{sec:model}

\begin{figure*}[htb]
    \centering
    \includegraphics[width=12cm]{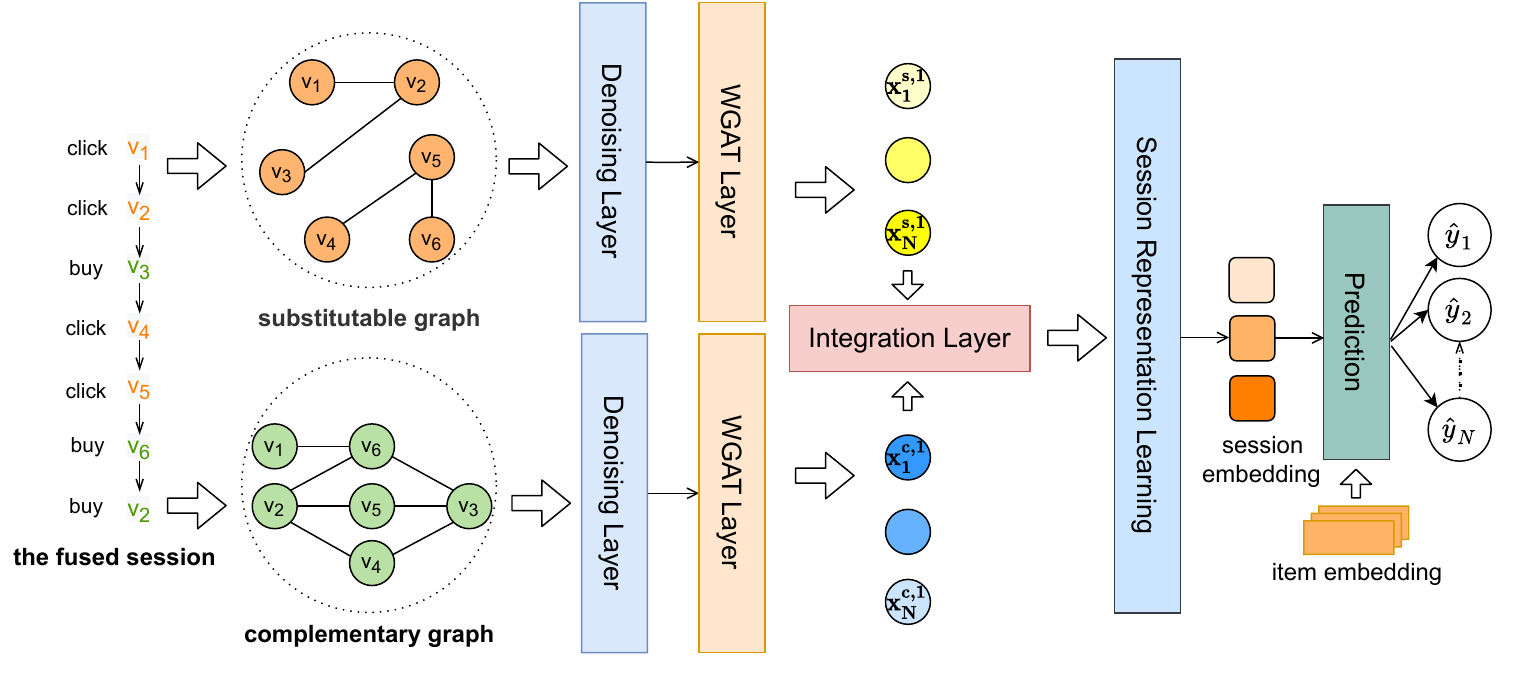}
    \caption{The overview of our proposed SCRM model.}
    \label{fig:model1}
\end{figure*}

Figure \ref{fig:model1} presents the architecture of SCRM, which is comprised of four main components: (1) \emph{graph construction}, and the design of $\mathcal{G}^s$ and $\mathcal{G}^c$ is detailed in the previous section. (2) \emph{item representation learning}, which removes task-irrelevant edges by denoising network and obtains the final item representation by incorporating both substitutable and complementary embeddings. (3) \emph{session representation learning}. It models user preference by aggregating the learned item representation. (4) \emph{prediction and loss function}, which strives to calculate the recommendation score $\hat{y}_j$ of each candidate item $v_j$. Loss function consists of three components: recommendation loss $L_r$, substitutable and complementary exclusivity loss $L_{ex}$ and semantic similarity loss $L_{se}$. We next present the four components in detail.

\subsection{Item Representation Learning}

\begin{figure}[htbp]
    \centering
    \includegraphics[width=9cm]{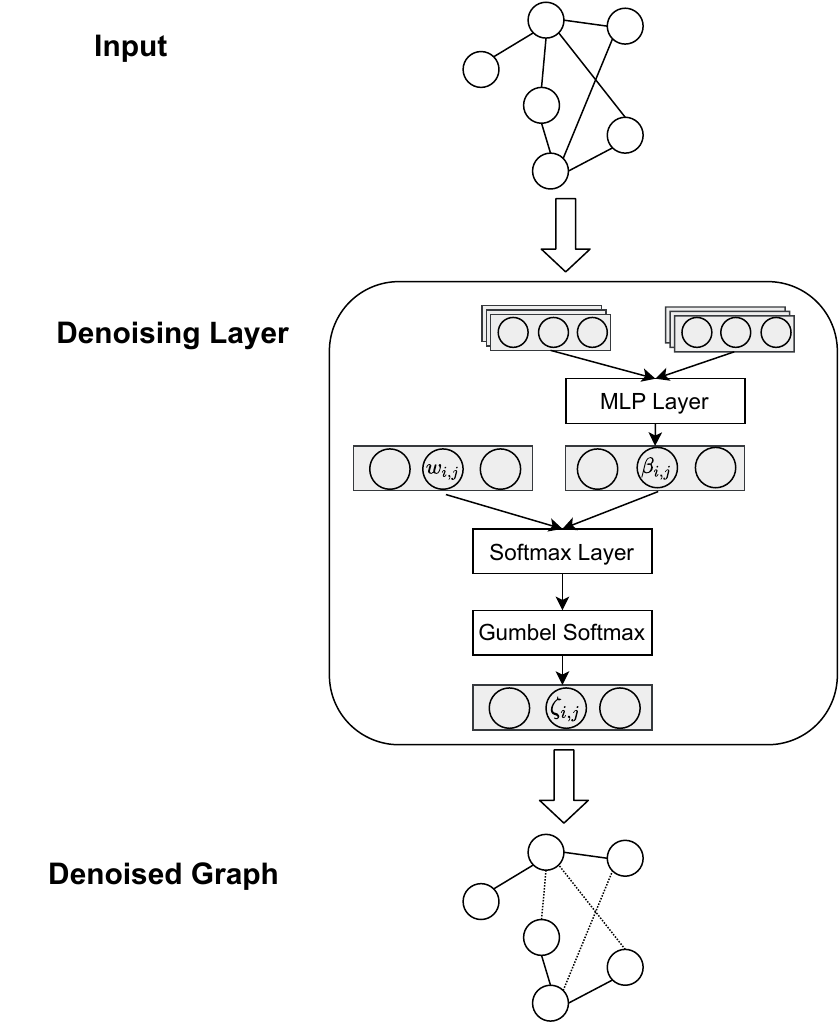}
    \caption{Denoising network.}
    \label{fig:DenoisingLayer}
\end{figure}
\subsubsection{Denoising layer}
As explained before, we establish the substitutable and complementary graphs by a set of intuitive rules that we summarize from data (prior knowledge), which inevitability introduce noise (although we have tried our best to cover many reliable connections) and also cannot describe all those intrinsic properties due to the uniqueness of different data. That is, the predefined rules are appropriate for most cases, however, not all extracted relationships in terms of these rules are 
exactly in the type of the corresponding relationship. For example, in the predefined rules, co-clicked products have substitutable relationship. However, in some cases, two products that are clicked adjacently may be complementary or uncorrelated.
Thus, we introduce two denoising networks to make the model automatically find the inherent substitutable and complementary connections and filter out the noisy ones from the constructed graphs, respectively. 

Towards the two graphs ($\mathcal{G}^s$ and $\mathcal{G}^c$), we use a similar method (Figure \ref{fig:DenoisingLayer}) adopted from \cite{zheng2020robust,luo2021learning} to filter out task-irrelevant edges by penalizing the number of edges with parameterized graphs and thus generate the corresponding updated graphs, respectively.
Take the substitutable graph as an example, we first denote $\mathbf{X}^0 \in \mathbb{R}^{N\times d_0}$ as the embedding matrix of items:
\begin{equation}
\small
\mathbf{X}^0=\text{nn.Embedding}(N, d_0)\label{eq:initialnnEmbedding}
\end{equation}

Later, we calculate the item embedding weight $\beta^s_{ij}$ with regard to each edge $E^s_{ij}$ in $\mathcal{G}^s$:
\begin{equation}\small
\begin{aligned}
\beta_{i,j}^s
&=MLP(\mathbf{x}^0_i,\mathbf{x}^0_j)\\
&=\text{softmax}(\mathbf{W}_{s,2}*(\mathbf{W}_{s,1}*[\mathbf{x}^0_i;\mathbf{x}^0_j]+\mathbf{\mu}_{s,1})+\mu_{s,2})
\end{aligned}\label{eq:denoising_weight1}
\end{equation}
where ; denotes concatenation operation. Besides, we use a two-layers MLP with learnable parameters $\mathbf{W}_{s,1} \in\mathbb{R}^{2d_0 \times 2d_0}$, $\mathbf{W}_{s,2} \in \mathbb{R}^{1 \times 2d_0}$, $\mathbf{\mu}_{s,1} \in\mathbb{R}^{2d_0 \times 1}$ and scalar $\mu_{s,2}$. The item embedding of item $v_i$, $\mathbf{x}_i^0=\mathbf{X}^0_{i,:}\in\mathbb{R}^{d_0 \times 1}$.
The embedding weight $\beta_{i,j}^s$ that we learned in the training process and the initial edge weight $w_{i,j}^s$ can both contain the substitutable information.
Therefore we add the embedding weight $\beta_{i,j}^s$ and initial edge weight $w_{i,j}^s$ to form the weight $z_{i,j}^s$ for denoising network: 
\begin{equation}\small
z_{i,j}^s=\beta_{i,j}^s+w_{i,j}^s\label{eq:denoising_weight}
\end{equation}

Thirdly, we compute the probability of each edge by employing a softmax function:
\begin{equation}\small
\pi_{i,j}^s=\frac{exp(z_{i,j}^s)}{\sum_{k \in N(i)}{exp(z_{i,k}^s)}}\label{eq:weight_softmax}
\end{equation}
where $N(i)$ is the set of connected items of $v_i$ in $\mathcal{G}^s$. We further use Gumbel-Softmax (to make the sampling process differentiable) to generate edge samples:
\begin{equation}\small
\zeta_{i,j}^s=\frac{exp(log(\pi_{i,j}^s+\epsilon)/\tau)}{\sum_{k \in N(i)}{exp(log(\pi_{i,k}^s+\epsilon)/\tau))}} \label{eq:gumble_softmax}
\end{equation}
where $q \thicksim \text{Uniform}(0,1)$ and $\epsilon=-log(-log(q))$. $\tau > 0$ is a temperature parameter.

Lastly, we reserve the top-$K_{\zeta}$ edges with the highest $\zeta$ values and delete the others. Thus, the learned $\zeta_{i,j}^s$ and $\zeta_{i,j}^c$ are the final weight for the corresponding edge of the substitutable and complementary graph, respectively.

\subsubsection{Item representation learning by integrating two graphs}
Given the updated substitutable and complementary graphs by the denoising layer, we adopt weighted graph attention network (WGAT) \cite{qiu2019rethinking} to obtain respective item representations, respectively. In particular, for example, for substitutable graph $\mathcal{G}^s$, we obtain the importance between $v_i$ and its neighbor $v_j$: 
\begin{equation}
\small
e^s_{ij}=\sigma(\mathbf{W}_{s,4}^T*[\mathbf{W}_{s,3}\mathbf{x}_i^0; \mathbf{W}_{s,3}\mathbf{x}_j^0; \zeta^s_{i,j}]) \label{eq:selfAttentionWeight}
\end{equation}
where $\sigma(.)$ is the Leaky ReLU function, $\mathbf{W}_{s,3}\in \mathbb{R}^{d_1 \times d_0}$ and $\mathbf{W}_{s,4}\in \mathbb{R}^{1 \times (2d_1+1)}$ are parameters.
We further adopt softmax function to normalize the $e^s_{ij}$:
\begin{equation}\small
\alpha^s_{ij}=\text{softmax}(e^s_{ij})=\frac {exp(e^s_{ij})}{\sum_{v_k\in N(i)} exp(e^s_{ik})}\label{eq:normalizedAttentionValue}
\end{equation}
Then, we linearly aggregate the information from neighbors to form item $v_i$'s embedding, $\mathbf{x}_i^{s,1}\in\mathbb{R}^{d_1 \times 1}$:
\begin{equation}\small
\mathbf{x}_{i}^{s,1}=\sigma (\sum\nolimits_{j\in N(i)} \alpha^s_{ij}\mathbf{W}_{s,5}\mathbf{x}_j^0) \label{eq:oneHead}
\end{equation}
where $\sigma(.)$ is the Leaky ReLU function and $\mathbf{W}_{s,5}\in \mathbb{R}^{d_1 \times d_0}$ is a trainable parameter matrix.
By using the Leaky ReLu function, the gradient can be calculated on the part of the input less than zero during backpropagation, thus avoiding the gradient direction sawtooth problem.
We average embedding from $K_m$ heads in multi-head attention mechanism to stabilize the training of the layers:
\begin{equation}\small
\mathbf{x_i^{s,1}}=\frac{1}{K_m} \sum\nolimits_{k\in[1,K_m]}\mathbf{x}_{i}^{k,s,1}\label{eq:finalItemEmbedding}
\end{equation}
where $\mathbf{x}_{i}^{k,s,1}$ is $v_i$'s embedding output by the $k$-th head. Besides, we can obtain complementary embedding $\mathbf{x_i^{c,1}}$ in the same way.  
\begin{figure}[htbp]
    \centering
    \includegraphics[width=10cm]{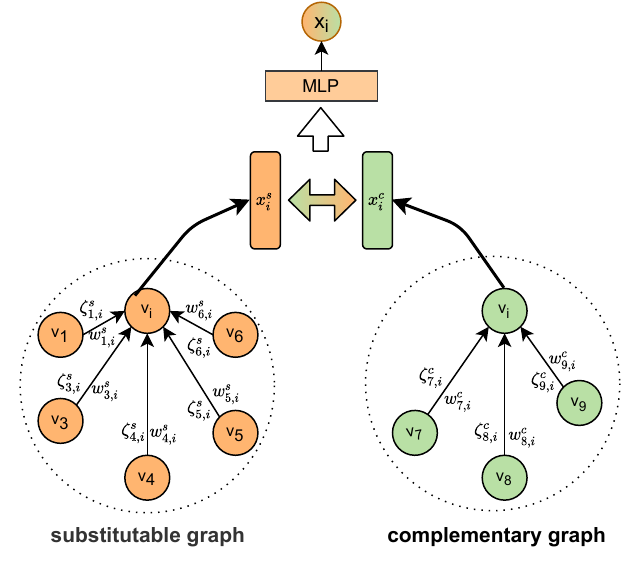}
    \caption{Integrating substitutable and complementary graphs.}
    \label{fig:interplay}
\end{figure}

An item's substitutable information (i.e., substitutes) can indicate its complementary information (i.e., complements), and vice versa. For example, Samsung phones and Huawei phones are substitutes, which are both under the category of mobile phones and have similar complementary products, such as chargers and earphones.
Hence, inspired by DecGCN \cite{liu2020decoupled}, item representation is learned by different subgraphs, which may influence each other mutually. 
After obtaining the embedding on each graph, we get the final substitutable (complementary) embedding by integrating the influence of $\mathbf{x_i^{c,1}}$ ($\mathbf{x_i^{s,1}}$) on $\mathbf{x_i^{s,1}}$ ($\mathbf{x_i^{c,1}}$) (Figure \ref{fig:interplay}):
\begin{equation}\small
\mathbf{x_i^{s}}=\sigma(\mathbf{x_i^{s,1}}+\theta_1 \mathbf{x_i^{c,1}})\label{eq:integrate_s}
\end{equation}
\begin{equation}\small
\mathbf{x_i^{c}}=\sigma(\mathbf{x_i^{c,1}}+\theta_2 \mathbf{x_i^{s,1}})\label{eq:integrate_c}
\end{equation}
where $\sigma(.)$ is Leaky RELU and $\theta_1$, $\theta_2$ are trainable parameters.

Finally, for each item $v_i\in\mathcal{V}$, we obtain its representation by  incorporating both substitutable embedding and complementary embedding with an MLP:
\begin{equation}\small
\mathbf{x_i}=MLP([\mathbf{x_i^{s}},\mathbf{x_i^{c}}])=\mathbf{W}_6*[\mathbf{x_i^{s}};\mathbf{x_i^{c}}]+\mathbf{\mu}_3 \label{eq:final_embedding}
\end{equation}
where $;$ denotes the concatenation operation, and $\mathbf{\mu}_3\in \mathbb{R}^{d_1 \times 1}$ and  $\mathbf{W}_{6} \in \mathbb{R}^{d_1\times 2d_1}$ are the learnable parameters.
\subsection{Session Representation Learning}
\label{sec:selearning}
We next present how to generate session representation. To represent the current session as an embedding, we plan to combine long-term preferences and current interests of the session.
In particular, we use the attention mechanism to compute a global level representation by aggregating item embeddings and treat the most recent behavior of the session as the current interests.

Consider the information in the session may have different importance. Given session $S_f$ ($S_f=\{v_1^S,\cdots,v_l^S\}$), we firstly use attention mechanism to compute the weighted factor $\alpha_k$ depicting the importance of $k$-th item ($v_k^S$) to $l$-th item (last item, $v_l^S$):
\begin{equation}\small
\alpha_k=\mathbf{q}^{T} \sigma (\mathbf{W}_{7} \mathbf{x}_{l} + \mathbf{W}_{8} \mathbf{x}_{k}+\mathbf{\mu_4})\label{eq:itemImportanceToSession}
\end{equation}
where $\sigma(.)$ is the Leaky RELU function. $\mathbf{q}\in\mathbb{R}^{d_1 \times 1}$, $\mathbf{\mu}_4\in \mathbb{R}^{d_1 \times 1}$, $\mathbf{W}_{7},\mathbf{W}_{8} \in \mathbb{R}^{d_1\times d_1}$ denote the learnable parameters. We then use these weighted factors to aggregate all item information to form the global session representation:
\begin{equation}\small
\mathbf{S}_g=\sum\nolimits_{k=1}^l \alpha_k \mathbf{x}_{k} \label{eq:globalSession}
\end{equation}

It is also important to explicitly consider users’ recent interests in session-based recommendation. Therefore we further project the concatenation to get the fused session representation by considering the information of the most recent behavior $\mathbf{x}_{l}$ and global session information $\mathbf{S}_g$:
\begin{equation}\small
\mathbf{S}=\mathbf{W}_{9} [\mathbf{x}_{l}; \mathbf{S}_g] \label{eq:sessionRepresentation}
\end{equation}
where $\mathbf{W}_{9}\in \mathbb{R}^{d_1\times 2d_1}$ is the projecting matrix.

\subsection{Prediction and Loss Function}
Based on the obtained session representation $\mathbf{S}$ and each item representation $\mathbf{x}_j$ of item $v_j$, we first compute the predicted probability of $v_j$.
Then, the predicted probability ($\hat{y}_j$) of the candidate item $v_j$ can be calculated as: 
\begin{equation}\small
\hat{y}_j=\text{softmax}(Score_j)=\frac{exp(\mathbf{S}^T \mathbf{x}_j)}{\sum\nolimits_{v_k\in\mathcal{V}}exp(\mathbf{S}^T \mathbf{x}_k)} \label{eq:recommendationScore}
\end{equation}

With the ground-truth $y_j$ (one-hot encoding) and predicted $\hat{y}_j$, our loss function is three-fold: recommendation loss ($L_r$), substitutable and complementary exclusivity loss ($L_{ex}$), and semantic similarity loss ($L_{se}$).
First, we adopt cross-entropy as the main \emph{recommendation loss} ($L_r$):
\begin{equation}\small
L_r=-\sum\nolimits_{j=1}^{N} y_j \log(\hat{y}_j)+(1-y_j)\log (1-\hat{y}_j) \label{eq:loss1}
\end{equation}

For the loss $L_{ex}$, considering that the substitutable and complementary relationships are mutually exclusive towards a product pair, i.e., it is comparatively rare to see that a product pair shows both highly substitutable and complementary relationships. Thus, we design $L_{ex}$ by maximizing the distance between substitutability and complementarity of $v_i$ and $v_j$:
\begin{equation}
L_{ex}=-\sum\nolimits_i{\sum\nolimits_j{ \sigma(((\mathbf{x}_i^{s})^T \mathbf{x}_j^{s}-(\mathbf{x}_i^{c})^T \mathbf{x}_j^{c}})^2)} \label{eq:loss2}
\end{equation}
where $v_i, v_j \in \mathcal{V}$ and $\sigma(x)=\frac{1}{1+e^{-x}}$.

Similar to \cite{zhao2017improving,zhang2019inferring}, the underlying assumption for \emph{semantic similarity loss function} ($L_{se}$) is that items that are viewed together ought to be more similar than items that are not. So for product $v_i$, given its substitute $v_j\in\mathcal{V}_i^{s}$ (the labelled set of $v_i$'s substitutes), complement $v_k\in\mathcal{V}_i^{c}$ (the set of $v_i$'s complements) and irrelevant product $v_t\in\mathcal{V}_i^{i}$ ($\mathcal{V}_i^{i}=\mathcal{V}-\mathcal{V}_i^{s}-\mathcal{V}_i^{c}-v_i$), we have following inequalities: 
\begin{equation}
\text{sim}(v_i,v_j)>\text{sim}(v_i,v_k)>\text{sim}(v_i,v_t)
\label{eq:loss31}
\end{equation}
It should be noted that we consider complementary products are more similar than unrelated products since there exists some correlation between complementary products, e.g., having the same usage scenarios or belonging to the same category, like milk and biscuits.
Accordingly, semantic similarity loss is defined as:
\begin{equation}
\begin{aligned}
L_{se} = &-\sum\nolimits_{v_i \in \mathcal{V}}{\sum\nolimits_{v_j \in \mathcal{V}_i^{s}}{\sum\nolimits_{v_k \in \mathcal{V}_i^{c}}{ \sigma(\mathbf{x}_i^T(\mathbf{x}_j-\mathbf{x}_k))}}}\\
         &-\sum\nolimits_{v_i \in \mathcal{V}}{\sum\nolimits_{v_k \in \mathcal{V}_i^{c}}{\sum\nolimits_{v_t \in \mathcal{V}_i^{i}}{ \sigma(\mathbf{x}_i^T(\mathbf{x}_k-\mathbf{x}_t))}}}
\end{aligned}
\label{eq:loss3}
\end{equation}
where $\sigma(x)=\frac{1}{1+e^{-x}}$.

Finally, the global loss function is given as follows:
\begin{equation}
L=L_{r}+ \gamma_1 L_{ex}+ \gamma_2 L_{se}
\label{eq:loss_last}
\end{equation}
where $\gamma_1$ and $\gamma_2$ are hyperparameters which control the trade off among different losses.

The Adam optimizer is exerted to optimize these parameters, where Algorithm \ref{alg:algorithm} summarizes the model procedure.

\begin{algorithm}[t]
\renewcommand{\algorithmicrequire}{\textbf{Input:}}
\renewcommand{\algorithmicensure}{\textbf{Output:}}
\caption{Overview of SCRM algorithm}
\label{alg:algorithm}
\begin{algorithmic}[1]
\Require Item set $\mathcal{V}$, The fused session set $\mathcal{S}_f$
\Ensure Top $K$ items according to each session
\State Construct substitutable graph $\mathcal{G}^s$ and complementary graph $ \mathcal{G}^c$;
\State Calculate initial weights $w_s$ and $w_c$;
\State Initialize the model parameter;
\For{epoch in epochs}
 \State Sample a batch $B$;
 \For{session $S_f$ in batch $B$}
  \State Update the substitutable and complementary graphs and corresponding weights through the denoising network;
  \State Learn two representations of items, $x^s$ and $x^c$, through different WGAT layer;
  \State Obtain item representations by fusing information on complementary graph and substitute graph by Equation \ref{eq:final_embedding};
 \State Learn session representations by fusing user long-term and short-term preferences through attention mechanism in Equation \ref{eq:sessionRepresentation};
 \State  Calculate the predicted probability $\hat{y}$ for each candidate items by Equation \ref{eq:recommendationScore};
 \State  Compute the loss $L_r$,$L_{ex}$ and $L_{se}$;
 \EndFor
  \State Use the gradient-based Adam optimizer to train the model
\EndFor

\end{algorithmic}
\end{algorithm}

\section{Experiments}
\label{sec:experiments}
\begin{table}[t]
\small\centering
\caption{Statistics of the datasets.}\label{tb:dataStatistics}
\begin{tabular}{@{}c@{}ccc}
\toprule
Dataset& Tmall& Yoochoose1/64 \\
\midrule
\#clicks \quad &86,629 &555,819  \\
\#purchases \quad &8,977  &20,770 \\
\#clicks/\#purchases \quad &9.65  &26.76 \\
\#items  \quad&10,157  &17,376  \\
\#train sessions  \quad &73,247  &330,733   \\
\#validation sessions  \quad&3,242   &57,091   \\
\#test sessions  \quad&3,242  &57,091  \\
average length \quad &6.02  &4.11  \\
\bottomrule
\end{tabular}
\end{table}

In this section, we conduct extensive experiments on two datasets to validate the effectiveness of SCRM, with the goal of answering the two research questions (RQs):

\begin{itemize}
\item \textbf{RQ1}: Can the proposed SCRM model learn substitutable and complementary relationships to improve session-based recommendation performance compared to other state-of-the-art approaches?
\item \textbf{RQ2}: Do different components of our SCRM model (e.g., loss function, denoising layer, substitutable graph, and complementary graph) improve the performance of session-based recommendation?
\end{itemize}

\subsection{Experimental Settings}

\subsubsection{Datasets}
We evaluate our model on two real-world datasets \emph{Tmall}\footnote{\url{tianchi.aliyun.com/dataset/dataDetail?dataId=42}.} and \emph{Yoochoose}\footnote{\url{www.kaggle.com/datasets/chadgostopp/recsys-challenge-2015}.}, which are commonly used in session-based recommendation\footnote{\emph{In our study, we adopt only Tmall and Yoochoose datasets from all the available public datatsets for session-based recommendation to evaluate our model mainly because they have both click and purchase session-based behaviors. For future work, we will take efforts to collect multi-behavior datasets to further verify t he effectiveness of our study (see Section VI).}}. In particular, %
\begin{itemize}
\item \emph{Tmall} is from IJCAI-15 contest and contains anonymized users' shopping logs in 6 months. 
We remove add-to-cart and add-to-favourite behaviors (interactions) on Tmall dataset. Regarding click and purchase behaviors, we firstly form the fused session $S_f$ in chronological order. Then, similar to \cite{wang2020global}, we filter out items with less than 5 interactions and sessions with lengths smaller than 2.
We set the most recent $3,242$ sessions as the test data, another $3,242$ sessions as the validation data, and the remaining ones for training. 
\item \emph{Yoochoose1/64} is used for RecSys Challenge 2015, and contains sequences that happened on an e-commerce site over six months. 
For the original dataset including click and purchase behavior types, we firstly form the fused session $S_f$ in chronological order. Then, similar to \cite{wu2019session,qiu2019rethinking}, we filter out items with less than 5 interactions and sessions with lengths smaller than 2.
We sort sessions with the increasing timestamp and take $57,091$ as the validation set, the last $57,091$ as the test set, the remaining and previous sessions as the training set. 
\end{itemize}
The statistics of the datasets are summarized in Table \ref{tb:dataStatistics}.
Besides, for both training and test data, regarding every session, we generate sequences by a splitting processing. That is, we finally could have more sessions besides the original one, e.g., $\{v_1^S,v_2^S\}$, $\{v_1^S,v_2^S, v_3^S\}$, $\{v_1^S,v_2^S, v_3^S,v_4^S\}$ for session $\mathcal{S}_f=\{v_1^S,v_2^S, v_3^S,v_4^S\}$.

\subsubsection{Baseline methods}
We compare our framework with two traditional methods (\textbf{POP} and \textbf{ItemKNN}), two RNN-based methods (\textbf{GRU4Rec} and \textbf{NARM}), and three state-of-the-art (SOTA) GNN-based methods (\textbf{SR-GNN}, \textbf{FGNN}, and \textbf{DHCN}) for session-based recommendation:
\begin{itemize}
\item \textbf{POP} recommends top-K frequent items in the training set;
\item \textbf{ItemKNN} \cite{sarwar2001item} recommends items that have the highest similarity (cosine similarity) with the last item of the session;
\item \textbf{GRU4Rec} \cite{hidasi2015session} utilizes gated recurrent units to process session data and adopts several modifications to classic recurrent neural networks such as a ranking loss function;
\item \textbf{NARM} \cite{li2017neural} employs recurrent neural network structures with vanilla attention to model the user’s main purpose and behavior for session-based recommendation problems;
\item \textbf{SR-GNN} \cite{wu2019session} applies a gated graph convolutional layer by combining long-term preferences and current interests of sessions to better predict users’ next actions;
\item \textbf{FGNN} \cite{qiu2019rethinking} proposes a weighted attention graph layer to learn item embeddings, and a Read-out function to obtain the session embeddings to represent the user’s preference;
\item \textbf{DHCN} \cite{xia2021self} introduces self-supervised learning to model the high-order correlations among users and items based on the hypergraph convolutional network.
\end{itemize}
Noted that aforementioned baseline models only use purchasing sequences or clicking sequences on datasets, whilst our model fuses two sequences in chronological order. Hence, for fair comparison we use the same session for baselines and our model.

\subsubsection{Evaluation metrics}
Following previous studies \cite{ren2019repeatnet,chen2020handling}, the performances are evaluated by three widely used metrics in session-based recommendation: Hit Ratio (\textbf{HR}@K), Mean Reciprocal Rank (\textbf{MRR}@K) and Normalized Discounted Cumulative Gain@$K$ (\textbf{NDCG}@K) and $K$ is set to $5$, $10$ and $20$ in our experiments. HR metric is an evaluation of unranked retrieval results, while the latter two are evaluations of ranked lists. Noted that for the three metrics, larger values indicate better performance.

\begin{itemize}
\item \textbf{HR}@$K$ denotes the hit ratio, i.e., the coverage rate of targeted predictions. This evaluation metric is the proportion of cases when the desired item is amongst the top-$K$ items in all test set. 
\begin{equation}
HR@K =\frac{1}{N} \sum_{i=0}^N hit(i)  \label{eq:hrMetric}
\end{equation}
where $N$ indicates the total number of accesses, that is, the actual number of clicks; $hit(i)$: If the recommendation system recommends item i, $hit(i)$ is 1, otherwise 0.

\item \textbf{MRR}@$K$ indicates the ranking accuracy based on the ranking position of the recommended items (hits), and a larger value means the ground-truth items are ranked in the top of the ranked recommendation lists. 
\begin{equation}
MRR@K =\frac{1}{N} \sum_{i=0}^N \frac{1}{p_i}  \label{eq:mrrMetric}
\end{equation}
where $p_i$ denotes the position of item i in the recommendation result. If item i does not appear, $p_i$ is $+\infty$.

\item \textbf{NDCG}@$K$ also rewards each hit based on its position in the ranked recommendation list.
\begin{equation}
DCG@K =\sum_{i=1}^K \frac{2^{r_i}-1}{\log_2(i+1)} \label{eq:dcg}
\end{equation}
\begin{equation}
NDCG@K =\frac{DCG@K}{IDCG}  \label{eq:ndcgMetric}
\end{equation}
where $r_i$ means the relevance of the recommendation result of position i. IDCG represents a list of the best recommended results returned by a user of the recommendation system.

\end{itemize}

\subsubsection{Hyper-parameters settings}
For SCRM, we apply one WGAT layer and all parameters are initialized using Gaussian distribution with a mean of $0$ and a standard deviation of $0.1$. The embedding size of each item $d_0$ and $d_1$ is $128$ on Tmall, $512$ on Yoochoose1/64. 
Besides, the initial learning rate is set to $0.001$ and $K_{\zeta}$ is set to $4$ for Tmall and Yoochoose 1/64 datasets. The initial temperature is set to $0.01$ and the weights of loss function, i.e., $\gamma_1$ and $\gamma_2$, are set to $0.2$ and $0.3$. We set the $L_2$ penalty to $1e-7$ and $1e-5$ on Tmall and Yoochoose 1/64, respectively. The batch size is $100$ on Tmall, $120$ on Yoochoose (the sample dataset from Yoochoose1/64 for RQ2), and $500$ on Yoochoose1/64.
For the baselines, we refer to their best parameters reported in the original papers as the similar datasets are explored and tune hyper-parameters to obtain the best performance on the rest of the datasets.

\subsection{Experimental Results}
Here, we present results to answer the aforementioned RQs.
\begin{table*}[htbp]
\renewcommand\tabcolsep{1pt}
\small\centering
 \caption{Performance of all comparison methods on two datasets. The best performance is boldfaced, and the runner-up is underlined. We compute the improvements that SCRM achieves relative to the best baseline. Besides, we adopt a paired t-test ($^{***}$ for p-value $\leq$.001) to have a statistical significance of pairwise differences of SCRM vs. the best baseline.}\label{tb:Results}
\resizebox{\textwidth}{35mm}{
\begin{tabular}{c|l|cc ccc cc c l }
\toprule

\textbf{Datasets} & \textbf{Metrics} & \textbf{POP} & \textbf{ItemKNN}  & \textbf{GRU4Rec} & \textbf{NARM}  & \textbf{SR-GNN} & \textbf{FGNN}  & \textbf{DHCN} & \textbf{SCRM}& \textbf{Improv.} \\
\hline
\multirow{9}{*}{Tmall}
& HR@5 &0.0045	&0.0823	&0.1354	&0.2593	&0.2896 &0.2603  &\underline{0.3947} &\textbf{0.5367} &35.98\%***  	\\
&  HR@10  &0.0061	&0.1017	&0.1614	&0.3269	&0.3600 &0.3041  &\underline{0.4876}  &\textbf{0.6131}  &25.74\%*** 	\\
 &  HR@20  &0.0096	&0.1176	&0.1851	&0.3997	&0.4181 &0.3461  &\underline{0.5697} &\textbf{0.6765}  &18.75\%***  	\\
 & MRR@5  &0.0016	&0.0527	&0.0832	&0.1779	&0.2089 &0.1845  &\underline{0.2845} &\textbf{0.3897} &36.98\%*** 	\\
&  MRR@10 &0.0018	&0.0554	&0.0867	&0.1868	&0.2182 &0.1904  &\underline{0.2956}  &\textbf{0.4000}  &35.32\%***  	\\
 &  MRR@20  &0.0020	&0.0565	&0.0883	&0.1920	&0.2222 &0.1934  &\underline{0.3010}  &\textbf{0.4045}  &34.39\%***  	\\
 &  NDCG@5  &0.0023	&0.0601	&0.0961	&0.1979	&0.2289 &0.2033  &\underline{0.3119} &\textbf{0.4265}  &36.74\%***  	\\
 & NDCG@10  &0.0028	&0.0664	&0.1046	&0.2197	&0.2515 &0.2176  &\underline{0.3385} &\textbf{0.4513}  &33.32\%***  	\\
 &  NDCG@20  &0.0037 &0.0704	&0.1106	&0.2382	&0.2661 &0.2283  &\underline{0.3581} &\textbf{0.4674}  &30.52\%***  	\\

\hline
\multirow{9}{*}{Yoochoose1/64} 
& HR@5   &0.0625 &0.2036  	&0.3845	 &0.4523	&\underline{0.4698} &0.4561  &0.4647  &\textbf{0.5016}  &6.77\%***  	\\
& HR@10  &0.0912 &0.2596    &0.5153	 &0.5825	&\underline{0.5978} &0.5866  &0.5810 &\textbf{0.6253} &4.60\%***  	\\
&HR@20   &0.1099 &0.3084  &0.6123	 &0.6877	&\underline{0.7028}	&0.6938  &0.6828    &\textbf{0.7299}  &3.86\%***  	\\
&MRR@5  &0.0284 &0.1249	&0.2024	 &0.2689	&\underline{0.2829} &0.2719  &0.2657  &\textbf{0.3052}  &7.88\%*** 	\\
&MRR@10 &0.0325 &0.1326 	&0.2201	 &0.2864	&\underline{0.3000} &0.2894  &0.2799  &\textbf{0.3219}  &7.30\%***  	\\
&MRR@20 &0.0338 &0.1361 	&0.2270	 &0.2938	&\underline{0.3074} &0.2969  &0.2863  &\textbf{0.3393}  &10.38\%*** 	\\
&NDCG@5   &0.0367 &0.1444 	&0.2474	 &0.3144	&\underline{0.3295} &0.3175  &0.3152 &\textbf{0.3542}  &7.50\%***  	\\
&NDCG@10  &0.0462 &0.1628    &0.2900	 &0.3566	&\underline{0.3708} &0.3599  &0.3496  &\textbf{0.3943}  &6.34\%***  	\\
&NDCG@20  &0.0509 &0.1753  &0.3146	 &0.3834	&\underline{0.3975} &0.3871  &0.3729 &\textbf{0.4210}  &5.91\%*** 	\\
\hline
\end{tabular}}
\end{table*}

\subsubsection{Overall Performance (RQ1).}
To demonstrate the overall performance of SCRM, we compare it with the aforementioned baseline methods. The results are presented in Table \ref{tb:Results}, from which we have the following observations. 
(1) The first two traditional methods achieve relatively poor performance on two datasets. In the traditional methods, POP has the worst performance, because it only recommends frequent top-K items. Comparing with POP, ItemKNN achieves better results among the traditional methods, which does not consider the chronological order of the items in the session.
(2) Compared with traditional methods, RNN-based and GNN-based methods have better performance for session-based recommendation. Between these RNN-based methods, NARM achieves better performance than GRU4REC. 
(3) GNN-based methods outperform RNN-based methods, but NARM only performs slightly worse, demonstrating that it is a rather competitive baseline for session-based recommendation. 
(4) SCRM achieves the best performance across all datasets in terms of all metrics, which proves the effectiveness of our proposed model. Specifically, SCRM achieves better performance than DHCN by 36.98\% on Tmall w.r.t. MRR@5. Furthermore, the improvements on Tmall are higher than those on Yoochoose. This is probably caused by that $\frac{\#purchase}{\#click}$ on Tmall is much larger (see Table \ref{tb:dataStatistics}), which makes it better to establish substitutable and complementary relationships.
\begin{table}[htb]
\small\centering
 \caption{Impact of loss function. The paired t-test is conducted between SCRM and SCRM-ex (* for p-value $\leq .05$, ** for p-value $\leq .01$, and *** for p-value $\leq .001$).}\label{tb:lossfunction}
\setlength{\tabcolsep}{0.8mm}{
\begin{tabular}{c|c|ccl }
\toprule
\textbf{Datasets} & \textbf{Metrics}   & \textbf{SCRM-ex}  & \textbf{SCRM-se} & \textbf{SCRM} \\
\hline
\multirow{4}{*}{Tmall}
& HR@5  &0.5066 &0.4981  & \textbf{0.5367***}\\
&  HR@10    &0.5781  &0.5688   & \textbf{0.6131***}\\
&  HR@20     &0.6393  &0.6318   & \textbf{0.6765***}\\
 & MRR@5      &0.3777  &0.3703   & \textbf{0.3897*}\\
 &  MRR@10    &0.3873  &0.3798   & \textbf{0.4000**}\\
 &  MRR@20    &0.3916  &0.3842  & \textbf{0.4045**}\\
&  NDCG@5    &0.4113  &0.4055   & \textbf{0.4265**}\\
 &  NDCG@10   &0.4331  &0.4252   & \textbf{0.4513**}\\
 &  NDCG@20   &0.4486  &0.4374   & \textbf{0.4674***}\\

\hline
\multirow{4}{*}{Yoochoose}
& HR@5  &0.5655  &0.5646  & \textbf{0.5764**}\\
&  HR@10    &0.6864  &0.6859  & \textbf{0.6972**}\\
&  HR@20     &0.7864  &0.7858   & \textbf{0.7945}\\
 & MRR@5      &0.3687  &0.3662   & \textbf{0.3780**}\\
 &  MRR@10    &0.3850  &0.3826   & \textbf{0.3943}\\
 &  MRR@20    &0.3920  &0.3897  & \textbf{0.4011*}\\
&  NDCG@5    &0.4177  &0.4156   & \textbf{0.4274*}\\
 &  NDCG@10   &0.4570  &0.4551   & \textbf{0.4667*}\\
 &  NDCG@20   &0.4823  &0.4806  & \textbf{0.4914}\\

\bottomrule
\end{tabular}}
\end{table}

 \begin{figure}[htb]
    \centering
	\footnotesize
	
	\begin{tikzpicture}
	\begin{groupplot}[group style={
		group name=myplot,
		group size= 2 by 2,  horizontal sep=1.5cm}, 
		height=5cm, width=5cm,
	ylabel style={yshift=-0.15cm},
	every tick label/.append style={font=\small}
	]

	\nextgroupplot[ybar=0.10,
	bar width=0.4em,
	ylabel={performance@10},
	xlabel={(a) Tmall},
	scaled ticks=false,
	ymin=0.3, ymax=0.9,
	enlarge x limits=0.4,
	symbolic x coords={HR,MRR,NDCG},
	ylabel style = {font=\small},
       xlabel style = {font=\small},
	legend style={font=\tiny},
	xtick=data,
	ytick={0.3,0.5,0.7,0.9},
	]
	\addplot[fill=orange] coordinates {
		(HR,0.6012) (MRR,0.3812) (NDCG,0.4341)
		};
	\addplot[fill=pink] coordinates {
    (HR,0.6131) (MRR,0.4000) (NDCG,0.4513)
  
		};
  \legend{SCRM-DL,SCRM}
    \nextgroupplot[ybar=0.10,
	bar width=0.4em,
	ylabel={performance@10},
	xlabel={(b) Yoochoose},
	scaled ticks=false,
	ymin=0.3, ymax=0.9,
	enlarge x limits=0.4,
	legend style={font=\tiny},
 	ylabel style = {font=\small},
        xlabel style = {font=\small},
	symbolic x coords={HR,MRR,NDCG},
	xtick=data,
	ytick={0.3,0.5,0.7,0.9},
	]
	\addplot[fill=orange] coordinates {
		(HR,0.6856) (MRR, 0.3810) (NDCG, 0.4537)};
	\addplot[fill=pink] coordinates {
		  (HR,0.6972) (MRR, 0.3943) (NDCG, 0.4667)};
    \legend{SCRM-DL,SCRM}
    
	\nextgroupplot[ybar=0.10,
	bar width=0.4em,
	ylabel={performance@20},
	xlabel={(a) Tmall},
	scaled ticks=false,
	ymin=0.3, ymax=0.9,
	enlarge x limits=0.4,
	symbolic x coords={HR,MRR,NDCG},
	ylabel style = {font=\small},
        xlabel style = {font=\small},
	legend style={font=\tiny},
	xtick=data,
	ytick={0.3,0.5,0.7,0.9},
	]
	\addplot[fill=orange] coordinates {
        (HR,0.6665) (MRR, 0.3858) (NDCG, 0.4507)};
	\addplot[fill=pink] coordinates {
  (HR,0.6765) (MRR,0.4045) (NDCG,0.4674)
		};
  \legend{SCRM-DL,SCRM}
    \nextgroupplot[ybar=0.10,
	bar width=0.4em,
	ylabel={performance@20},
	xlabel={(b) Yoochoose},
	scaled ticks=false,
	ymin=0.3, ymax=0.9,
	enlarge x limits=0.4,
	legend style={font=\tiny},
 	ylabel style = {font=\small},
        xlabel style = {font=\small},
	symbolic x coords={HR,MRR,NDCG},
	xtick=data,
	ytick={0.3,0.5,0.7,0.9},
	]
	\addplot[fill=orange] coordinates {
		(HR,0.7875) (MRR, 0.3882) (NDCG, 0.4796)};
	\addplot[fill=pink] coordinates {
		(HR,0.7945) (MRR, 0.4011) (NDCG, 0.4914)};
  
    \legend{SCRM-DL,SCRM}
    
	\end{groupplot}
    \end{tikzpicture}
     \caption{The impact of denoising layer ($K=10, 20$).}
    \label{fig:denoisingLayer}
\end{figure}
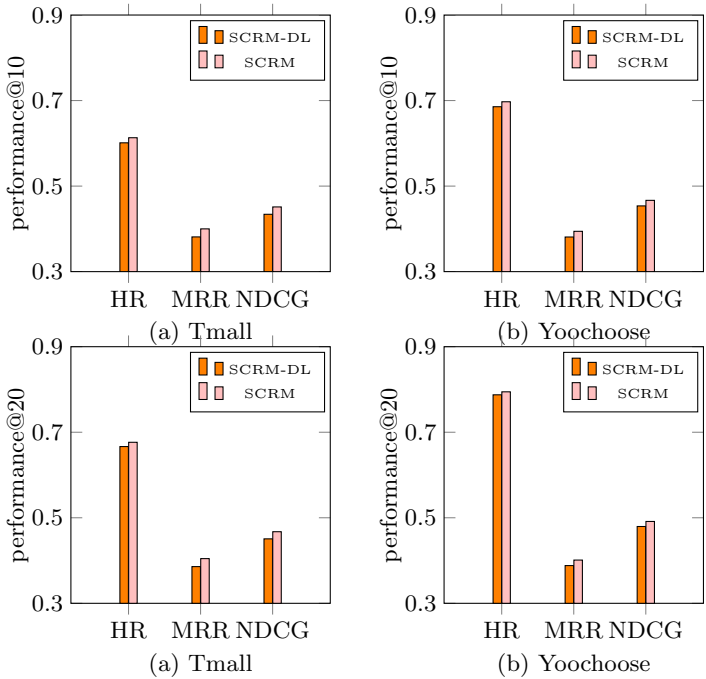

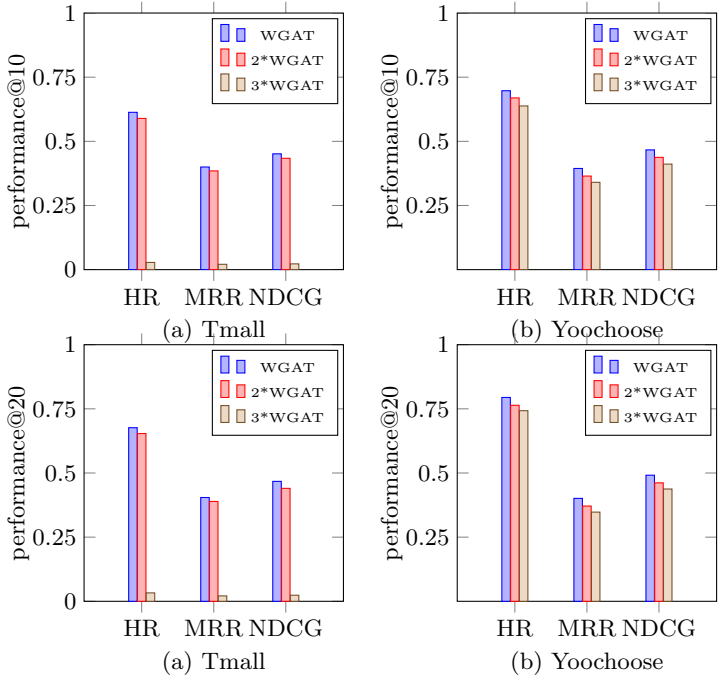
\begin{figure}[htb]
    \centering
	\footnotesize
	
	\begin{tikzpicture}
	\begin{groupplot}[group style={
		group name=myplot,
		group size= 2 by 2,  horizontal sep=1.5cm}, 
		height=5cm, width=5cm,
	ylabel style={yshift=-0.15cm},
	every tick label/.append style={font=\small}
	]

	\nextgroupplot[ybar=0.10,
	bar width=0.4em,
	ylabel={performance@10},
	xlabel={(a) Tmall},
	scaled ticks=false,
	ymin=0, ymax=1.0,
	enlarge x limits=0.4,
	symbolic x coords={HR,MRR,NDCG},
	ylabel style = {font=\small},
        xlabel style = {font=\small},
	legend style={font=\tiny},
	xtick=data,
	ytick={0,0.25,0.5,0.75,1.0},
	]
	\addplot coordinates {
		(HR,0.6131) (MRR,0.4000) (NDCG,0.4513)
		};
	\addplot coordinates {
		(HR,0.5894) (MRR,0.3847) (NDCG,0.4339)
		};

	\addplot coordinates {
		 (HR,0.0280) (MRR,0.0208) (NDCG,0.0225)
		};
  \legend{WGAT,2*WGAT,3*WGAT}
    \nextgroupplot[ybar=0.10,
	bar width=0.4em,
	ylabel={performance@10},
	xlabel={(b) Yoochoose},
	scaled ticks=false,
	ymin=0, ymax=1,
	enlarge x limits=0.4,
	legend style={font=\tiny},
 	ylabel style = {font=\small},
        xlabel style = {font=\small},
	symbolic x coords={HR,MRR,NDCG},
	xtick=data,
	ytick={0.25,0.5,0.75,1},
	]
 	\addplot coordinates {
		  (HR,0.6972) (MRR, 0.3943) (NDCG, 0.4667)};
	\addplot coordinates {
		(HR,0.6692) (MRR, 0.3647) (NDCG, 0.4376)};
	\addplot coordinates {
		(HR,0.6379) (MRR, 0.3403) (NDCG, 0.4113)};
    \legend{WGAT,2*WGAT,3*WGAT}
  
	\nextgroupplot[ybar=0.10,
	bar width=0.4em,
	ylabel={performance@20},
	xlabel={(a) Tmall},
	scaled ticks=false,
	ymin=0, ymax=1.0,
	enlarge x limits=0.4,
	symbolic x coords={HR,MRR,NDCG},
	ylabel style = {font=\small},
	legend style={font=\tiny},
        xlabel style = {font=\small},
	xtick=data,
	ytick={0,0.25,0.5,0.75,1.0},
	]
	\addplot coordinates {
		(HR,0.6765) (MRR,0.4045) (NDCG,0.4674)
		};

	\addplot coordinates {
		 (HR,0.6537) (MRR,0.3892) (NDCG,0.4402)
		};
	\addplot coordinates {
		 (HR,0.0328) (MRR,0.0211) (NDCG,0.0237)
		};
  \legend{WGAT,2*WGAT,3*WGAT}
  
    \nextgroupplot[ybar=0.10,
	bar width=0.4em,
	ylabel={performance@20},
	xlabel={(b) Yoochoose},
	scaled ticks=false,
	ymin=0, ymax=1,
	enlarge x limits=0.4,
	legend style={font=\tiny},
 	ylabel style = {font=\small},
        xlabel style = {font=\small},
	symbolic x coords={HR,MRR,NDCG},
	xtick=data,
	ytick={0.25,0.5,0.75,1},
	]
	\addplot coordinates {
		(HR,0.7945) (MRR, 0.4011) (NDCG, 0.4914)};
	\addplot coordinates {
		  (HR,0.7639) (MRR, 0.3713) (NDCG, 0.4616)};
    	\addplot coordinates {
		  (HR,0.7428) (MRR, 0.3476) (NDCG, 0.4376)};
    \legend{WGAT,2*WGAT,3*WGAT}

	\end{groupplot}
    \end{tikzpicture}
    \caption{The impact of the different number of WGAT layers ($K=10,20$).}
    \label{fig:differentWGAT}
\end{figure}

\begin{table}[htb]
\small\centering
  \caption{Impact of substitutable and complementary graphs.}\label{tb:subcomGraph}
\setlength{\tabcolsep}{0.8mm}{
\begin{tabular}{c|c|ccc }
\toprule
\textbf{Datasets} & \textbf{Metrics}   & \textbf{SCRM-S}  & \textbf{SCRM-C} & \textbf{SCRM} \\
\hline
\multirow{4}{*}{Tmall}
& HR@5  &0.4810  &0.4637  & \textbf{0.5367}\\
&  HR@10    &0.5487  &0.5204  & \textbf{0.6131}\\
&  HR@20     &0.6121  &0.5821  & \textbf{0.6765}\\
 & MRR@5      &0.3748  &0.3362  & \textbf{0.3897}\\
 &  MRR@10    &0.3838  &0.3438  & \textbf{0.4000}\\
 &  MRR@20    &0.3882  &0.3481  & \textbf{0.4045}\\
&  NDCG@5    &0.4013  &0.3755  & \textbf{0.4265}\\
 &  NDCG@10   &0.4232  &0.3939  & \textbf{0.4513}\\
 &  NDCG@20   &0.4392  &0.4095  & \textbf{0.4674}\\

\hline
\multirow{4}{*}{Yoochoose}
& HR@5  &0.5349  &0.5255  & \textbf{0.5764}\\
&  HR@10    &0.6583  &0.6481  & \textbf{0.6972}\\
&  HR@20     &0.7633   &0.7535  & \textbf{0.7945}\\
 & MRR@5      &0.3421 &0.3397  & \textbf{0.3780}\\
 &  MRR@10    &0.3587  &0.3562  & \textbf{0.3943}\\
 &  MRR@20    &0.3661  &0.3636  & \textbf{0.4011}\\
&  NDCG@5    &0.3901  &0.3859  & \textbf{0.4274}\\
 &  NDCG@10   &0.4302  &0.4257  & \textbf{0.4667}\\
 &  NDCG@20   &0.4568 &0.4525 & \textbf{0.4914}\\

\bottomrule
\end{tabular}}
\end{table}

 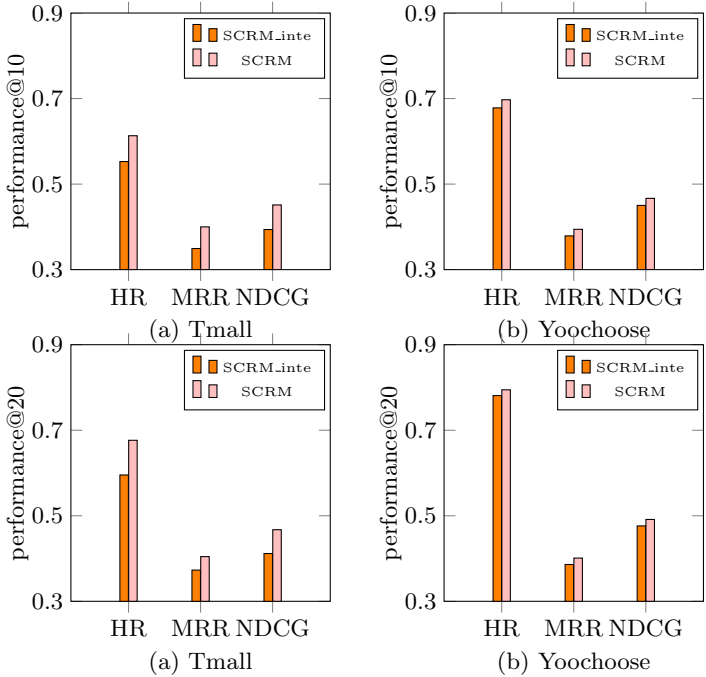
\begin{figure}[htb]
    \centering
	\footnotesize
	
	\begin{tikzpicture}
	\begin{groupplot}[group style={
		group name=myplot,
		group size= 2 by 2,  horizontal sep=1.5cm}, 
		height=5cm, width=5cm,
	ylabel style={yshift=-0.15cm},
	every tick label/.append style={font=\small}
	]

	\nextgroupplot[ybar=0.10,
	bar width=0.4em,
	ylabel={performance@10},
	xlabel={(a) Tmall},
	scaled ticks=false,
	ymin=0.3, ymax=0.9,
	enlarge x limits=0.4,
	symbolic x coords={HR,MRR,NDCG},
	ylabel style = {font=\small},
       xlabel style = {font=\small},
	legend style={font=\tiny},
	xtick=data,
	ytick={0.3,0.5,0.7,0.9},
	]
	\addplot[fill=orange] coordinates {
		(HR,0.5527) (MRR,0.3494) (NDCG,0.3938)
		};
	\addplot[fill=pink] coordinates {
    (HR,0.6131) (MRR,0.4000) (NDCG,0.4513)
  
		};
  \legend{SCRM\_inte,SCRM}
    \nextgroupplot[ybar=0.10,
	bar width=0.4em,
	ylabel={performance@10},
	xlabel={(b) Yoochoose},
	scaled ticks=false,
	ymin=0.3, ymax=0.9,
	enlarge x limits=0.4,
	legend style={font=\tiny},
 	ylabel style = {font=\small},
        xlabel style = {font=\small},
	symbolic x coords={HR,MRR,NDCG},
	xtick=data,
	ytick={0.3,0.5,0.7,0.9},
	]
	\addplot[fill=orange] coordinates {
		(HR,0.6782) (MRR, 0.3789) (NDCG, 0.4502)};
	\addplot[fill=pink] coordinates {
		  (HR,0.6972) (MRR, 0.3943) (NDCG, 0.4667)};
    \legend{SCRM\_inte,SCRM}

	\nextgroupplot[ybar=0.10,
	bar width=0.4em,
	ylabel={performance@20},
	xlabel={(a) Tmall},
	scaled ticks=false,
	ymin=0.3, ymax=0.9,
	enlarge x limits=0.4,
	symbolic x coords={HR,MRR,NDCG},
	ylabel style = {font=\small},
        xlabel style = {font=\small},
	legend style={font=\tiny},
	xtick=data,
	ytick={0.3,0.5,0.7,0.9},
	]
	\addplot[fill=orange] coordinates {
        (HR,0.5954) (MRR, 0.3730) (NDCG, 0.4117)};
	\addplot[fill=pink] coordinates {
  (HR,0.6765) (MRR,0.4045) (NDCG,0.4674)
		};
  \legend{SCRM\_inte,SCRM}
    \nextgroupplot[ybar=0.10,
	bar width=0.4em,
	ylabel={performance@20},
	xlabel={(b) Yoochoose},
	scaled ticks=false,
	ymin=0.3, ymax=0.9,
	enlarge x limits=0.4,
	legend style={font=\tiny},
 	ylabel style = {font=\small},
        xlabel style = {font=\small},
	symbolic x coords={HR,MRR,NDCG},
	xtick=data,
	ytick={0.3,0.5,0.7,0.9},
	]
	\addplot[fill=orange] coordinates {
		(HR,0.7810) (MRR, 0.3861) (NDCG, 0.4763)};
	\addplot[fill=pink] coordinates {
		(HR,0.7945) (MRR, 0.4011) (NDCG, 0.4914)};
  
    \legend{SCRM\_inte,SCRM}

	\end{groupplot}
    \end{tikzpicture}
    \caption{The interplay of substitutability and complementarity ($K=10, 20$).}
    \label{fig:Interplay_of_SC}
\end{figure}

 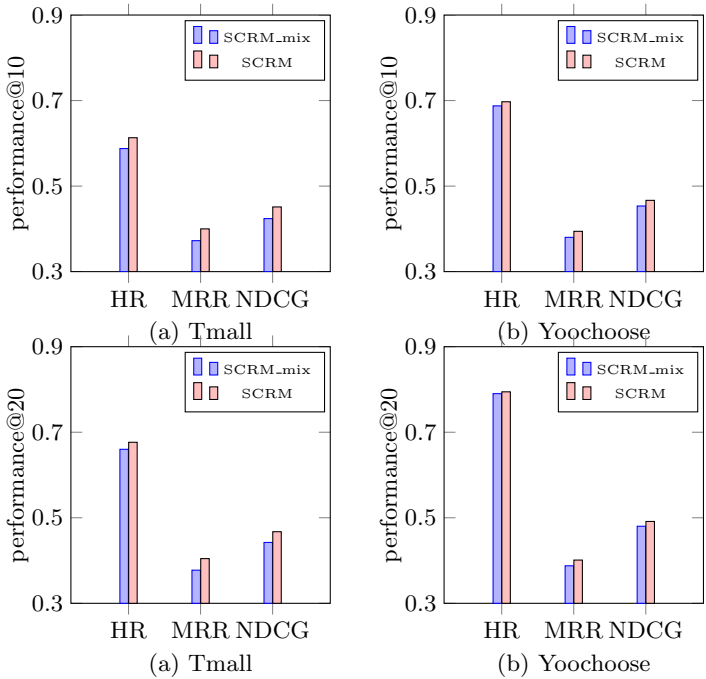
\begin{figure}[htb]
    \centering
	\footnotesize
	
	\begin{tikzpicture}
	\begin{groupplot}[group style={
		group name=myplot,
		group size= 2 by 2,  horizontal sep=1.5cm}, 
		height=5cm, width=5cm,
	ylabel style={yshift=-0.15cm},
	every tick label/.append style={font=\small}
	]

	\nextgroupplot[ybar=0.10,
	bar width=0.4em,
	ylabel={performance@10},
	xlabel={(a) Tmall},
	scaled ticks=false,
	ymin=0.3, ymax=0.9,
	enlarge x limits=0.4,
	symbolic x coords={HR,MRR,NDCG},
	ylabel style = {font=\small},
        xlabel style = {font=\small},
	legend style={font=\tiny},
	xtick=data,
	ytick={0.3,0.5,0.7,0.9},
	]
	\addplot coordinates {
		(HR,0.5878) (MRR,0.3724) (NDCG,0.4240) 
		};
	\addplot[fill=pink] coordinates {
		(HR,0.6131) (MRR,0.4000) (NDCG,0.4513)
		};
  \legend{SCRM\_mix,SCRM}
    \nextgroupplot[ybar=0.10,
	bar width=0.4em,
	ylabel={performance@10},
	xlabel={(b) Yoochoose},
	scaled ticks=false,
	ymin=0.3, ymax=0.9,
	enlarge x limits=0.4,
        ylabel style = {font=\small},
        xlabel style = {font=\small},
	legend style={font=\tiny},
	symbolic x coords={HR,MRR,NDCG},
	xtick=data,
	ytick={0.3,0.5,0.7,0.9},
	]
	\addplot coordinates {
		(HR,0.6876) (MRR, 0.3802) (NDCG, 0.4535)};
	\addplot[fill=pink] coordinates {
		  (HR,0.6972) (MRR, 0.3943) (NDCG, 0.4667)};
    \legend{SCRM\_mix,SCRM}

	\nextgroupplot[ybar=0.10,
	bar width=0.4em,
	ylabel={performance@20},
	xlabel={(a) Tmall},
	scaled ticks=false,
	ymin=0.3, ymax=0.9,
	enlarge x limits=0.4,
	symbolic x coords={HR,MRR,NDCG},
	ylabel style = {font=\small},
        xlabel style = {font=\small},
	legend style={font=\tiny},
	xtick=data,
	ytick={0.3,0.5,0.7,0.9},
	]
	\addplot coordinates {
		(HR,0.6601) (MRR,0.3775) (NDCG,0.4423)
		};
	\addplot[fill=pink] coordinates {
  (HR,0.6765) (MRR,0.4045) (NDCG,0.4674) 
		};
  \legend{SCRM\_mix,SCRM}
    \nextgroupplot[ybar=0.10,
	bar width=0.4em,
	ylabel={performance@20},
	xlabel={(b) Yoochoose},
	scaled ticks=false,
	ymin=0.3, ymax=0.9,
	enlarge x limits=0.4,
	ylabel style = {font=\small},
        xlabel style = {font=\small},
	legend style={font=\tiny},
	symbolic x coords={HR,MRR,NDCG},
	xtick=data,
	ytick={0.3,0.5,0.7,0.9},
	]
	\addplot coordinates {
		(HR,0.7902) (MRR, 0.3877) (NDCG, 0.4802)};
	\addplot[fill=pink] coordinates {
		(HR,0.7945) (MRR, 0.4011) (NDCG, 0.4914)};
    \legend{SCRM\_mix,SCRM}

	\end{groupplot}
    \end{tikzpicture}
 \caption{Impact of the separation of two graphs ($K=10, 20$).}
     \label{fig:separationGraphs}
\end{figure}

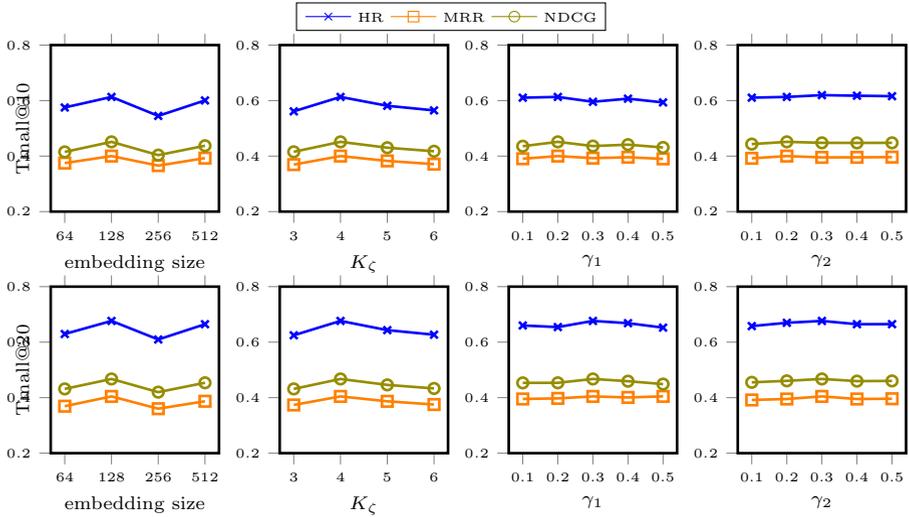
\begin{figure*}[tbp]
    \centering
	\footnotesize
	\begin{tikzpicture}
      \matrix[
          matrix of nodes,
          draw,
          inner sep=0.2em,
          ampersand replacement=\&,
          font=\tiny,
          anchor=south
        ]
        { 
		\ref{plots:HR} HR
		\ref{plots:MRR} MRR
		\ref{plots:NDCG} NDCG\\
          };
    \end{tikzpicture}\\
\begin{tikzpicture}
	\begin{groupplot}[group style={
		group name=myplot,
		group size= 4 by 2,  horizontal sep=0.8cm}, 
		height=3.8cm, width=3.8cm,
	ylabel style={yshift=-0.5cm},
	tick align=outside,
	every tick label/.append style={font=\tiny}
	]

\nextgroupplot[
ylabel=Tmall@10,
xlabel=embedding size,
ymin=0.2,
ymax=0.8,
ytick={0.2,0.4,0.6,0.8},
scaled ticks=false,
symbolic x coords={64,128,256,512},
xtick=data,
line width=1pt,
ylabel style = {font=\footnotesize},
xlabel style = {font=\footnotesize},
legend style={font=\tiny},
ytick pos=left
]

\addplot[color=blue,mark=x] coordinates {
(64,0.5751 )  (128,0.6131 )  (256, 0.5448)  (512,0.6005)
};\label{plots:HR}
\addplot[color=orange,mark=square] coordinates {
(64, 0.3748)  (128, 0.4)  (256, 0.3657)  (512,0.3926)
};\label{plots:MRR}
\addplot[color=olive,mark=o] coordinates {
(64, 0.415)  (128,0.4513)  (256, 0.4035)  (512,0.4373)
};\label{plots:NDCG}

\nextgroupplot[
xlabel=$K_{\zeta}$,
ymin=0.2,
ymax=0.8,
ytick={0.2,0.4,0.6,0.8},
scaled ticks=false,
symbolic x coords={3,4,5,6},
xtick=data,
line width=1pt,
xlabel style = {font=\footnotesize},
legend style={font=\tiny},
ytick pos=left
]

\addplot[color=blue,mark=x] coordinates {
(3,0.5612)  (4,0.6131)  (5, 0.5812)  (6,0.5646)
};
\addplot[color=orange,mark=square] coordinates {
(3, 0.3693)  (4,0.4)  (5, 0.3826)  (6,0.3709)
};
\addplot[color=olive,mark=o] coordinates {
(3, 0.4154)  (4,0.4513)  (5,0.4302)  (6,0.4174)
};

\nextgroupplot[
xlabel=$\gamma_1$,
ymin=0.2,
ymax=0.8,
ytick={0.2,0.4,0.6,0.8},
scaled ticks=false,
symbolic x coords={0.1,0.2,0.3,0.4,0.5},
xtick=data,
line width=1pt,
xlabel style = {font=\footnotesize},
legend style={font=\tiny},
ytick pos=left
]

\addplot[color=blue,mark=x] coordinates {
(0.1,0.6104)  (0.2,0.6131 )  (0.3, 0.5956)  (0.4,0.6069) (0.5,0.5934)
};
\addplot[color=orange,mark=square] coordinates {
(0.1, 0.3904)  (0.2, 0.4)  (0.3, 0.3926)  (0.4,0.3958) (0.5,0.3899)
};
\addplot[color=olive,mark=o] coordinates {
(0.1, 0.4356)  (0.2,0.4513)  (0.3, 0.4361)  (0.4,0.4412) (0.5,0.4311)
};

\nextgroupplot[
xlabel=$\gamma_2$,
ymin=0.2,
ymax=0.8,
ytick={0.2,0.4,0.6,0.8},
scaled ticks=false,
symbolic x coords={0.1,0.2,0.3,0.4,0.5},
xtick=data,
line width=1pt,
xlabel style = {font=\footnotesize},
legend style={font=\tiny},
ytick pos=left
]

\addplot[color=blue,mark=x] coordinates {
(0.1, 0.6105)  (0.2,0.6131 )  (0.3, 0.6196)  (0.4,0.6175) (0.5,0.6158)
};
\addplot[color=orange,mark=square] coordinates {
(0.1, 0.3920) (0.2, 0.4)  (0.3, 0.3953)  (0.4,0.3955) (0.5,0.3963)
};
\addplot[color=olive,mark=o] coordinates {
(0.1, 0.4432)  (0.2,0.4513)  (0.3, 0.4478)  (0.4,0.4475) (0.5,0.4478)
};

\nextgroupplot[
ylabel=Tmall@20,
xlabel=embedding size,
ymin=0.2,
ymax=0.8,
ytick={0.2,0.4,0.6,0.8},
scaled ticks=false,
symbolic x coords={64,128,256,512},
xtick=data,
line width=1pt,
ylabel style = {font=\footnotesize},
xlabel style = {font=\footnotesize},
legend style={font=\tiny},
ytick pos=left
]
\addplot[color=blue,mark=x] coordinates {
(64,0.6294)  (128,0.6765 )  (256, 0.6096)  (512,0.6646)
};
\addplot[color=orange,mark=square] coordinates {
(64, 0.3693)  (128, 0.4045)  (256, 0.3602)  (512,0.3871)
};
\addplot[color=olive,mark=o] coordinates {
(64, 0.4315)  (128,0.4674)  (256, 0.4199)  (512,0.4536)
};

\nextgroupplot[
xlabel=$K_{\zeta}$,
ymin=0.2,
ymax=0.8,
ytick={0.2,0.4,0.6,0.8},
scaled ticks=false,
symbolic x coords={3,4,5,6},
xtick=data,
line width=1pt,
xlabel style = {font=\footnotesize},
legend style={font=\tiny},
ytick pos=left
]

\addplot[color=blue,mark=x] coordinates {
(3,0.6248)  (4,0.6765 )  (5, 0.6433)  (6,0.6269)
};
\addplot[color=orange,mark=square] coordinates {
(3, 0.3737)  (4, 0.4045)  (5, 0.3869)  (6,0.3753)
};
\addplot[color=olive,mark=o] coordinates {
(3, 0.4315)  (4,0.4674 )  (5, 0.446)  (6,0.4332)
};

\nextgroupplot[
xlabel=$\gamma_1$,
ymin=0.2,
ymax=0.8,
ytick={0.2,0.4,0.6,0.8},
scaled ticks=false,
symbolic x coords={0.1,0.2,0.3,0.4,0.5},
xtick=data,
line width=1pt,
xlabel style = {font=\footnotesize},
legend style={font=\tiny},
ytick pos=left
]

\addplot[color=blue,mark=x] coordinates {
(0.1,0.6601)  (0.2,0.6543)  (0.3, 0.6765) (0.4,0.6684) (0.5,0.6524)
};
\addplot[color=orange,mark=square] coordinates {
(0.1,0.3953)  (0.2,0.3974 )   (0.3, 0.4045)  (0.4,0.4008) (0.5,0.4047)
};
\addplot[color=olive,mark=o] coordinates {
(0.1,0.4533 )  (0.2,0.4535 )   (0.3, 0.4674)  (0.4,0.4593) (0.5,0.4489)
};

\nextgroupplot[
xlabel=$\gamma_2$,
ymin=0.2,
ymax=0.8,
ytick={0.2,0.4,0.6,0.8},
scaled ticks=false,
symbolic x coords={0.1,0.2,0.3,0.4,0.5},
xtick=data,
line width=1pt,
xlabel style = {font=\footnotesize},
legend style={font=\tiny},
ytick pos=left
]

\addplot[color=blue,mark=x] coordinates {
(0.1,0.6578 )  (0.2,0.6698 )  (0.3, 0.6765 ) (0.4,0.6647) (0.5,0.6649)
};
\addplot[color=orange,mark=square] coordinates {
(0.1,0.3917 )  (0.2,0.3952 )   (0.3, 0.4045)  (0.4,0.3952) (0.5,0.3961)
};
\addplot[color=olive,mark=o] coordinates {
(0.1,0.4552)  (0.2,0.4606)   (0.3, 0.4674)  (0.4,0.4596) (0.5,0.4603)
};

\end{groupplot}

\end{tikzpicture}
 \caption{Impact of different hyper-parameters on Tmall dataset (K=$10,20$).}
\label{fig:hyper_tmall} 
\end{figure*}

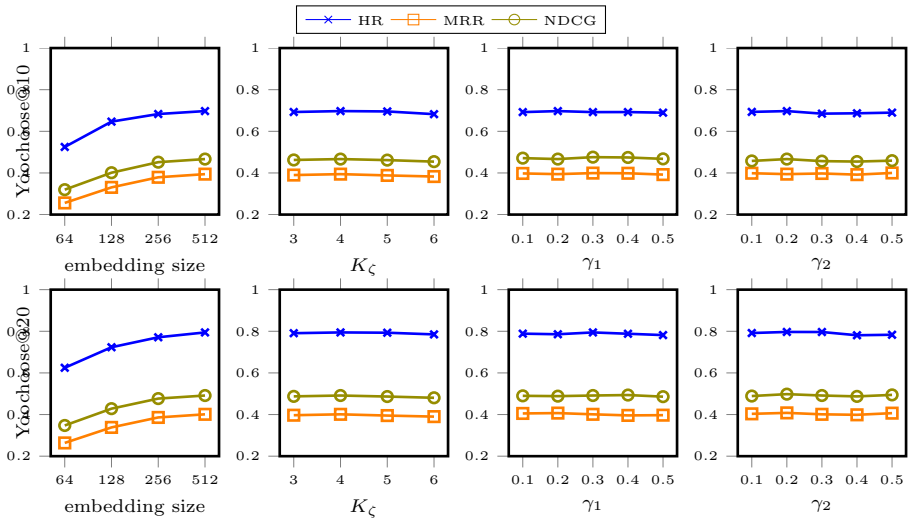
\begin{figure*}[tbp]
    \centering
	\footnotesize
	\begin{tikzpicture}
      \matrix[
          matrix of nodes,
          draw,
          inner sep=0.2em,
          ampersand replacement=\&,
          font=\tiny,
          anchor=south
        ]
        { 
		\ref{plots:HR} HR
		\ref{plots:MRR} MRR
		\ref{plots:NDCG} NDCG\\
          };
    \end{tikzpicture}\\
\begin{tikzpicture}
	\begin{groupplot}[group style={
		group name=myplot,
		group size= 4 by 2,  horizontal sep=0.8cm}, 
		height=3.8cm, width=3.8cm,
	ylabel style={yshift=-0.5cm},
	tick align=outside,
	every tick label/.append style={font=\tiny}
	]

\nextgroupplot[
ylabel=Yoochoose@10,
xlabel=embedding size,
ymin=0.2,
ymax=1.0,
ytick={0.2,0.4,0.6,0.8,1.0},
scaled ticks=false,
symbolic x coords={64,128,256,512},
xtick=data,
line width=1pt,
ylabel style = {font=\footnotesize},
xlabel style = {font=\footnotesize},
legend style={font=\tiny},
ytick pos=left
]

\addplot[color=blue,mark=x] coordinates {
(64,0.5249 )  (128, 0.6466)  (256, 0.6831)  (512,0.6972)
};\label{plots:HR}
\addplot[color=orange,mark=square] coordinates {
(64, 0.2560)  (128, 0.3307)  (256, 0.3792)  (512,0.3943)
};\label{plots:MRR}
\addplot[color=olive,mark=o] coordinates {
(64, 0.3197)  (128,0.4012)  (256,0.4518 )  (512,0.4667)
};\label{plots:NDCG}

\nextgroupplot[
xlabel=$K_{\zeta}$,
ymin=0.2,
ymax=1.0,
ytick={0.2,0.4,0.6,0.8,1.0},
scaled ticks=false,
symbolic x coords={3,4,5,6},
xtick=data,
line width=1pt,
xlabel style = {font=\footnotesize},
legend style={font=\tiny},
ytick pos=left
]

\addplot[color=blue,mark=x] coordinates {
(3,0.6928)  (4,0.6972)  (5, 0.6955)  (6,0.6824)
};
\addplot[color=orange,mark=square] coordinates {
(3, 0.3899)  (4,0.3943)  (5, 0.3882)  (6,0.3828)
};
\addplot[color=olive,mark=o] coordinates {
(3, 0.4624)  (4,0.4667)  (5,0.4617)  (6,0.4544)
};

\nextgroupplot[
xlabel=$\gamma_1$,
ymin=0.2,
ymax=1.0,
ytick={0.2,0.4,0.6,0.8,1.0},
scaled ticks=false,
symbolic x coords={0.1,0.2,0.3,0.4,0.5},
xtick=data,
line width=1pt,
xlabel style = {font=\footnotesize},
legend style={font=\tiny},
ytick pos=left
]

\addplot[color=blue,mark=x] coordinates {
(0.1,0.6923)  (0.2,0.6972 )  (0.3, 0.6923)  (0.4,0.6925) (0.5,0.6896)
};
\addplot[color=orange,mark=square] coordinates {
(0.1,0.3977)  (0.2, 0.3943)  (0.3, 0.3993)  (0.4,0.3985) (0.5,0.3921)
};
\addplot[color=olive,mark=o] coordinates {
(0.1,0.4712 )  (0.2,0.4667)  (0.3, 0.4759)  (0.4,0.4744) (0.5,0.4677)
};

\nextgroupplot[
xlabel=$\gamma_2$,
ymin=0.2,
ymax=1.0,
scaled ticks=false,
symbolic x coords={0.1,0.2,0.3,0.4,0.5},
xtick=data,
line width=1pt,
xlabel style = {font=\footnotesize},
legend style={font=\tiny},
ytick pos=left
]

\addplot[color=blue,mark=x] coordinates {
(0.1,0.6933)  (0.2,0.6972 )  (0.3, 0.6849)  (0.4,0.6866) (0.5,0.6898)
};
\addplot[color=orange,mark=square] coordinates {
(0.1,0.3987)  (0.2, 0.3943)  (0.3, 0.3975)  (0.4,0.3919) (0.5,0.4000)
};
\addplot[color=olive,mark=o] coordinates {
(0.1,0. 4577)  (0.2,0.4667)  (0.3, 0.4568)  (0.4,0.4547) (0.5,0.4591)
};

\nextgroupplot[
ylabel=Yoochoose@20,
xlabel=embedding size,
ymin=0.2,
ymax=1.0,
ytick={0.2,0.4,0.6,0.8,1.0},
scaled ticks=false,
symbolic x coords={64,128,256,512},
xtick=data,
line width=1pt,
ylabel style = {font=\footnotesize},
xlabel style = {font=\footnotesize},
legend style={font=\tiny},
ytick pos=left
]
\addplot[color=blue,mark=x] coordinates {
(64, 0.6247)  (128,0.7233 )  (256, 0.7712)  (512,0.7945)
};\label{plots:HR}
\addplot[color=orange,mark=square] coordinates {
(64, 0.2637)  (128,0.3382 )  (256,0.3861 )  (512,0.4011)
};\label{plots:MRR}
\addplot[color=olive,mark=o] coordinates {
(64, 0.3475)  (128,0.4283)  (256, 0.4767)  (512,0.4914)
};\label{plots:NDCG}

\nextgroupplot[
xlabel=$K_{\zeta}$,
ymin=0.2,
ymax=1.0,
ytick={0.2,0.4,0.6,0.8,1.0},
scaled ticks=false,
symbolic x coords={3,4,5,6},
xtick=data,
line width=1pt,
xlabel style = {font=\footnotesize},
legend style={font=\tiny},
ytick pos=left
]

\addplot[color=blue,mark=x] coordinates {
(3,0.7908)  (4,0.7945 )  (5, 0.7931)  (6,0.785)
};
\addplot[color=orange,mark=square] coordinates {
(3, 0.3968)  (4, 0.4011)  (5, 0.3951)  (6,0.3901)
};
\addplot[color=olive,mark=o] coordinates {
(3, 0.4873)  (4,0.4914 )  (5, 0.4865)  (6,0.4805)
};

\nextgroupplot[
xlabel=$\gamma_1$,
ymin=0.2,
ymax=1.0,
ytick={0.2,0.4,0.6,0.8,1.0},
scaled ticks=false,
symbolic x coords={0.1,0.2,0.3,0.4,0.5},
xtick=data,
line width=1pt,
xlabel style = {font=\footnotesize},
legend style={font=\tiny},
ytick pos=left
]

\addplot[color=blue,mark=x] coordinates {
(0.1,0.7886)  (0.2,0.7857)  (0.3, 0.7945 )  (0.4,0.7881) (0.5,0.7817)
};
\addplot[color=orange,mark=square] coordinates {
(0.1,0.4053)  (0.2, 0.4070)  (0.3, 0.4011)  (0.4,0.3959) (0.5,0.3970)
};
\addplot[color=olive,mark=o] coordinates {
(0.1,0. 4899)  (0.2,0.4884)  (0.3, 0.4914)  (0.4,0.4937) (0.5,0.4859)
};

\nextgroupplot[
xlabel=$\gamma_2$,
ymin=0.2,
ymax=1.0,
scaled ticks=false,
symbolic x coords={0.1,0.2,0.3,0.4,0.5},
xtick=data,
line width=1pt,
xlabel style = {font=\footnotesize},
legend style={font=\tiny},
ytick pos=left
]

\addplot[color=blue,mark=x] coordinates {
(0.1,0.7916)  (0.2,0.7968 )  (0.3, 0.7965 )  (0.4,0.7809) (0.5,0.7833)
};
\addplot[color=orange,mark=square] coordinates {
(0.1,0.4034)  (0.2, 0.4080)  (0.3, 0.4011)  (0.4,0.3987) (0.5,0.4067)
};
\addplot[color=olive,mark=o] coordinates {
(0.1,0. 4889)  (0.2,0.4980)  (0.3, 0.4914)  (0.4,0.4872) (0.5,0.4944)
};

\end{groupplot}

\end{tikzpicture}
 \caption{Impact of different hyper-parameters on Yoochoose dataset (K=$10,20$).}
\label{fig:hyper_Yoochoose} 
\end{figure*}

\subsubsection{Impact of loss function (RQ2).}
To investigate the effectiveness of each module in loss function, we develop two variants of SCRM: 
\begin{itemize}
\item SCRM-ex, which removes substitutable and complementary exclusivity loss function ($L_{ex}$) in Equation \ref{eq:loss_last}; 
\item SCRM-se, which drops semantic similarity loss function ($L_{se}$) in Equation \ref{eq:loss_last}.
\end{itemize}

The results are presented in Table \ref{tb:lossfunction}, where we conduct a paired t-test ($^{*}$ for p-value $\leq$.05, $^{**}$ for p-value $\leq$.01, and $^{***}$ for p-value $\leq$.001) between SCRM and SCRM-ex\footnote{The significance test is compared between SCRM with SCRM-ex since SCRM-se performs consistently worse than SCRM-ex, indicating the larger contribution of $L_{se}$ than that of $L_{ex}$. In this case, the significance results in Table \ref{tb:lossfunction} can be at least applied to the pair of SCRM and SCRM-se.}. Since the Yoochoose 1/64 dataset is too large, we only use the most recent 1/8 fractions of the training sessions in this and the following experiments. As we can see, each component contributes to the final performance. This demonstrates that substitutable and complementary exclusivity loss and semantic similarity loss are beneficial to learning item representations, which is consistent with our hypothesis. Regarding substitutable and complementary exclusivity loss, this implies that the substitutable and complementary relationships are exclusive towards a product pair. As for the semantic similarity loss, this is because substitutable products ought to be more similar than items that are not. Moreover, complementary products have the same usage scenarios, and there exists some correlation between them, such as milk and biscuits belonging to the food category, whilst unrelated products do not have much relevance.

\subsubsection{Impact of denoising layer (RQ2).}
The denoising layer is used to rule out the noise in each graph and make the model better learn substitutable and complementary relationships. To evaluate the effectiveness of denoising layer, we design a model variant SCRM-DL, which does not consider denoising layer in \emph{Item Representation Learning}. Figure \ref{fig:denoisingLayer} shows the performance of SCRM-DL and SCRM, which confirms that denoising network can filter the noise and make the model perform better. This result occurs because not all relationships conform to the predefined rules.  Therefore, in our paper, we first design some rules that are appropriate for the majority of cases. Then, we use denoising networks to automatically filter out the noisy (false) relationships from the constructed graphs. By doing this, we try to keep the most possibly substitutable and complementary relationship in the graph, respectively.

\subsubsection{The impact of the different number of WGAT layers. (RQ2).}
To further explore the impact of the different number of WGAT layers, we consider model variants by varying the number of WGAT layers. Accordingly, we get three variants, namely, WGAT, 2*WGAT, and 3*WGAT (SCRM with one-layer WGAT, two-layer WGAT, and three-layer WGAT, respectively). The comparative results are summarized in Figure \ref{fig:differentWGAT}, which shows that one layer WGAT consistently performs better across all scenarios, indicating the possible noises involved in high-order relationships. This might be due to there being second-order relationships in the complementary graph. Therefore two layer WGAT means high-order item relationships in graphs.

\subsubsection{Impact of substitutable graph vs. complementary graph (RQ2).}
We further conduct experiments to explore the effectiveness of each graph. Specifically, we develop two variants of SCRM: SCRM-S merely considers complementary graph, while SCRM-C only uses substitutable graph. Noted that substitutable and complementary exclusivity loss, and semantic similarity loss contain both elements. Hence, the loss function of two variants only involves the main recommendation loss. Besides, the item embeddings in Equation \ref{eq:final_embedding} only adopt the representation of each graph separately. For example, in SCRM-C model, $x_i=x_i^s=x_i^{s,1}$. Table \ref{tb:subcomGraph} shows the comparative results between different models, which clearly show that SCRM can achieve better performance than only using one graph, and thus validates the importance of each graph.

\subsubsection{The interplay of substitutability and complementarity in Item Representation Learning (RQ2).}
In order to better learn substitute and complement representations, we integrate the influence of $\mathbf{x_i^{c,1}}$ ($\mathbf{x_i^{s,1}}$) on $\mathbf{x_i^{s,1}}$ ($\mathbf{x_i^{c,1}}$) by Equations \ref{eq:integrate_s} and  \ref{eq:integrate_c}. Here, we conduct analysis of interplay of the two relationships, which aims at validating the effectiveness of our design. Specially, we design the variant SCRM\_inte, which does not integrate the inter-relationship. Figure \ref{fig:Interplay_of_SC} shows the comparisons between SCRM and SCRM\_inte, which demonstrates that SCRM performs much better than SCRM\_inte in terms of all the three metrics. This verifies the effectiveness of our design.

\subsubsection{Impact of the separation of two graphs (RQ2).}
In this paper, we construct two graphs (substitutable and complementary graphs) along with two WGATs to learn item representation separately. 
Here, we conduct experiments to analyze the effect of this way of distinguishing substitutable and complementary graphs. Specifically, we introduce one variant: SCRM\_mix, which firstly fuses edges from these two graphs, and then adopts WGAT to learn item representations from the fused graph. Similarly, the loss function of this variant only involves the main recommendation loss. Figure \ref{fig:separationGraphs} shows the performance of SCRM and its variant SCRM\_mix on two datasets. Firstly, by distinguishing substitutable graph from complementary one, SCRM outperforms its variant SCRM\_mix across all scenarios. 
Secondly, we also find that SCRM\_mix achieves better performance compared with other state-of-the-art models, which shows that the effectiveness of our design on identifying item relationships.

\subsubsection{Impact of hyper-parameters (RQ2).}
Here, we study how the embedding size, $K_{\zeta}$ and hyper-parameters of loss function ($\gamma_1, \gamma_2$) affect our performance. We range the embedding size of SCRM within $\{64,128,256,512\}$, $K_{\zeta}$ in \{3,4,5,6\} and $\gamma$ in $\{0.1,0.2,0.3,0.4,0.5\}$. 
Figures \ref{fig:hyper_tmall} and \ref{fig:hyper_Yoochoose} illustrate the results in terms of different hyper-parameters. As for the embedding size, when it equals $128$, SCRM has the best performance on Tmall, whilst with the increase of embedding size, the performance improves on Yoochoose. Besides, SCRM is not very sensitive on both datasets for $K_{\zeta}$, $\gamma_1$ and $\gamma_2$.

\section{Conclusion}
\label{sec:conclusions}
In this paper, we proposed a novel method denoted as SCRM to explore substitutable and complementary relationships from multi-behavior data (i.e., purchase and click) for session-based recommendation. Our work clearly differed from the previous studies which either considered sessions of different behavior types separately, or ignored to explore substitutable and complementary relationships for better SR.
In our work, we constructed substitutable and complementary graphs to better leverage the sequential dependencies between different behavior types and thus capture these two relationships between items. We further introduced denoising layer to rule out the noise in constructed graphs and designed loss functions for effective session-based recommendation.
Extensive experiments on two real-world datasets demonstrated the effectiveness of our model and its components. 
 
For future work, we consider adding more behavior types to our model. We also strive to design other models to provide explicit and suitable explainability for substitution and complementary relationships. Besides, we will discuss the influence of data sparsity on session-based recommendation and try to design models to solve this problem.
Moreover, we will take efforts to collect multi-behavior datasets to further verify the effectiveness of our approach.

\backmatter

\section*{Declarations}
\bmhead{Conflict of interest}
The authors have no relevant financial or non-financial interests to disclose.

\bmhead{Funding}
We greatly acknowledge the support of National Natural Science Foundation of China (Grant No. 72371148 and 72192832), the Shanghai Rising-Star Program (Grant No. 23QA1403100), and the Natural Science Foundation of Shanghai (Grant No. 21ZR1421900).

\bibliography{references}


\begin{thebibliography}{48}
\ifx \bisbn   \undefined \def \bisbn  #1{ISBN #1}\fi
\ifx \binits  \undefined \def \binits#1{#1}\fi
\ifx \bauthor  \undefined \def \bauthor#1{#1}\fi
\ifx \batitle  \undefined \def \batitle#1{#1}\fi
\ifx \bjtitle  \undefined \def \bjtitle#1{#1}\fi
\ifx \bvolume  \undefined \def \bvolume#1{\textbf{#1}}\fi
\ifx \byear  \undefined \def \byear#1{#1}\fi
\ifx \bissue  \undefined \def \bissue#1{#1}\fi
\ifx \bfpage  \undefined \def \bfpage#1{#1}\fi
\ifx \blpage  \undefined \def \blpage #1{#1}\fi
\ifx \burl  \undefined \def \burl#1{\textsf{#1}}\fi
\ifx \doiurl  \undefined \def \doiurl#1{\url{https://doi.org/#1}}\fi
\ifx \betal  \undefined \def \betal{\textit{et al.}}\fi
\ifx \binstitute  \undefined \def \binstitute#1{#1}\fi
\ifx \binstitutionaled  \undefined \def \binstitutionaled#1{#1}\fi
\ifx \bctitle  \undefined \def \bctitle#1{#1}\fi
\ifx \beditor  \undefined \def \beditor#1{#1}\fi
\ifx \bpublisher  \undefined \def \bpublisher#1{#1}\fi
\ifx \bbtitle  \undefined \def \bbtitle#1{#1}\fi
\ifx \bedition  \undefined \def \bedition#1{#1}\fi
\ifx \bseriesno  \undefined \def \bseriesno#1{#1}\fi
\ifx \blocation  \undefined \def \blocation#1{#1}\fi
\ifx \bsertitle  \undefined \def \bsertitle#1{#1}\fi
\ifx \bsnm \undefined \def \bsnm#1{#1}\fi
\ifx \bsuffix \undefined \def \bsuffix#1{#1}\fi
\ifx \bparticle \undefined \def \bparticle#1{#1}\fi
\ifx \barticle \undefined \def \barticle#1{#1}\fi
\bibcommenthead
\ifx \bconfdate \undefined \def \bconfdate #1{#1}\fi
\ifx \botherref \undefined \def \botherref #1{#1}\fi
\ifx \url \undefined \def \url#1{\textsf{#1}}\fi
\ifx \bchapter \undefined \def \bchapter#1{#1}\fi
\ifx \bbook \undefined \def \bbook#1{#1}\fi
\ifx \bcomment \undefined \def \bcomment#1{#1}\fi
\ifx \oauthor \undefined \def \oauthor#1{#1}\fi
\ifx \citeauthoryear \undefined \def \citeauthoryear#1{#1}\fi
\ifx \endbibitem  \undefined \def \endbibitem {}\fi
\ifx \bconflocation  \undefined \def \bconflocation#1{#1}\fi
\ifx \arxivurl  \undefined \def \arxivurl#1{\textsf{#1}}\fi
\csname PreBibitemsHook\endcsname

\bibitem{rendle2010factorizing}
\begin{bchapter}
\bauthor{\bsnm{Rendle}, \binits{S.}},
\bauthor{\bsnm{Freudenthaler}, \binits{C.}},
\bauthor{\bsnm{Schmidt-Thieme}, \binits{L.}}:
\bctitle{Factorizing personalized markov chains for next-basket recommendation}.
In: \bbtitle{Proceedings of the 19th International Conference on World Wide Web},
pp. \bfpage{811}--\blpage{820}
(\byear{2010})
\end{bchapter}
\endbibitem

\bibitem{hidasi2018recurrent}
\begin{bchapter}
\bauthor{\bsnm{Hidasi}, \binits{B.}},
\bauthor{\bsnm{Karatzoglou}, \binits{A.}}:
\bctitle{Recurrent neural networks with top-k gains for session-based recommendations}.
In: \bbtitle{Proceedings of the 27th ACM International Conference on Information and Knowledge Management},
pp. \bfpage{843}--\blpage{852}
(\byear{2018})
\end{bchapter}
\endbibitem

\bibitem{zhang2021personal}
\begin{barticle}
\bauthor{\bsnm{Zhang}, \binits{X.}},
\bauthor{\bsnm{Zhou}, \binits{Y.}},
\bauthor{\bsnm{Wang}, \binits{J.}},
\bauthor{\bsnm{Lu}, \binits{X.}}:
\batitle{Personal interest attention graph neural networks for session-based recommendation}.
\bjtitle{Entropy}
\bvolume{23}(\bissue{11}),
\bfpage{1500}
(\byear{2021})
\end{barticle}
\endbibitem

\bibitem{kang2018self}
\begin{bchapter}
\bauthor{\bsnm{Kang}, \binits{W.-C.}},
\bauthor{\bsnm{McAuley}, \binits{J.}}:
\bctitle{Self-attentive sequential recommendation}.
In: \bbtitle{IEEE International Conference on Data Mining},
pp. \bfpage{197}--\blpage{206}
(\byear{2018}).
\bcomment{IEEE}
\end{bchapter}
\endbibitem

\bibitem{kipf2016semi}
\begin{bchapter}
\bauthor{\bsnm{Kipf}, \binits{T.N.}},
\bauthor{\bsnm{Welling}, \binits{M.}}:
\bctitle{Semi-supervised classification with graph convolutional networks}.
In: \bbtitle{Proceedings of the 5th International Conference on Learning Representations}
(\byear{2017})
\end{bchapter}
\endbibitem

\bibitem{velivckovic2017graph}
\begin{bchapter}
\bauthor{\bsnm{Veli{\v{c}}kovi{\'c}}, \binits{P.}},
\bauthor{\bsnm{Cucurull}, \binits{G.}},
\bauthor{\bsnm{Casanova}, \binits{A.}},
\bauthor{\bsnm{Romero}, \binits{A.}},
\bauthor{\bsnm{Lio}, \binits{P.}},
\bauthor{\bsnm{Bengio}, \binits{Y.}}:
\bctitle{Graph attention networks}.
In: \bbtitle{Proceedings of the 6th International Conference on Learning Representations}
(\byear{2018})
\end{bchapter}
\endbibitem

\bibitem{fang2020deep}
\begin{barticle}
\bauthor{\bsnm{Fang}, \binits{H.}},
\bauthor{\bsnm{Zhang}, \binits{D.}},
\bauthor{\bsnm{Shu}, \binits{Y.}},
\bauthor{\bsnm{Guo}, \binits{G.}}:
\batitle{Deep learning for sequential recommendation: Algorithms, influential factors, and evaluations}.
\bjtitle{ACM Transactions on Information Systems (TOIS)}
\bvolume{39}(\bissue{1}),
\bfpage{1}--\blpage{42}
(\byear{2020})
\end{barticle}
\endbibitem

\bibitem{mas1995microeconomic}
\begin{bbook}
\bauthor{\bsnm{Mas-Colell}, \binits{A.}},
\bauthor{\bsnm{Whinston}, \binits{M.D.}},
\bauthor{\bsnm{Green}, \binits{J.R.}}, \betal:
\bbtitle{Microeconomic Theory}
vol. \bseriesno{1}.
\bpublisher{Oxford University Press},
\blocation{New York}
(\byear{1995})
\end{bbook}
\endbibitem

\bibitem{varian2014intermediate}
\begin{bbook}
\bauthor{\bsnm{Varian}, \binits{H.R.}}:
\bbtitle{Intermediate Microeconomics: A Modern Approach: Ninth International Student Edition}.
\bpublisher{WW Norton \& Company},
\blocation{New York}
(\byear{2014})
\end{bbook}
\endbibitem

\bibitem{zhang2018quality}
\begin{bchapter}
\bauthor{\bsnm{Zhang}, \binits{Y.}},
\bauthor{\bsnm{Lu}, \binits{H.}},
\bauthor{\bsnm{Niu}, \binits{W.}},
\bauthor{\bsnm{Caverlee}, \binits{J.}}:
\bctitle{Quality-aware neural complementary item recommendation}.
In: \bbtitle{Proceedings of the 12th ACM Conference on Recommender Systems},
pp. \bfpage{77}--\blpage{85}
(\byear{2018})
\end{bchapter}
\endbibitem

\bibitem{wang2018path}
\begin{bchapter}
\bauthor{\bsnm{Wang}, \binits{Z.}},
\bauthor{\bsnm{Jiang}, \binits{Z.}},
\bauthor{\bsnm{Ren}, \binits{Z.}},
\bauthor{\bsnm{Tang}, \binits{J.}},
\bauthor{\bsnm{Yin}, \binits{D.}}:
\bctitle{A path-constrained framework for discriminating substitutable and complementary products in e-commerce}.
In: \bbtitle{Proceedings of the 11th ACM International Conference on Web Search and Data Mining},
pp. \bfpage{619}--\blpage{627}
(\byear{2018})
\end{bchapter}
\endbibitem

\bibitem{mcauley2015inferring}
\begin{bchapter}
\bauthor{\bsnm{McAuley}, \binits{J.}},
\bauthor{\bsnm{Pandey}, \binits{R.}},
\bauthor{\bsnm{Leskovec}, \binits{J.}}:
\bctitle{Inferring networks of substitutable and complementary products}.
In: \bbtitle{Proceedings of the 21th ACM SIGKDD International Conference on Knowledge Discovery and Data Mining},
pp. \bfpage{785}--\blpage{794}
(\byear{2015})
\end{bchapter}
\endbibitem

\bibitem{rakesh2019linked}
\begin{bchapter}
\bauthor{\bsnm{Rakesh}, \binits{V.}},
\bauthor{\bsnm{Wang}, \binits{S.}},
\bauthor{\bsnm{Shu}, \binits{K.}},
\bauthor{\bsnm{Liu}, \binits{H.}}:
\bctitle{Linked variational autoencoders for inferring substitutable and supplementary items}.
In: \bbtitle{Proceedings of the 12th ACM International Conference on Web Search and Data Mining},
pp. \bfpage{438}--\blpage{446}
(\byear{2019})
\end{bchapter}
\endbibitem

\bibitem{liu2020decoupled}
\begin{bchapter}
\bauthor{\bsnm{Liu}, \binits{Y.}},
\bauthor{\bsnm{Gu}, \binits{Y.}},
\bauthor{\bsnm{Ding}, \binits{Z.}},
\bauthor{\bsnm{Gao}, \binits{J.}},
\bauthor{\bsnm{Guo}, \binits{Z.}},
\bauthor{\bsnm{Bao}, \binits{Y.}},
\bauthor{\bsnm{Yan}, \binits{W.}}:
\bctitle{Decoupled graph convolution network for inferring substitutable and complementary items}.
In: \bbtitle{Proceedings of the 29th ACM International Conference on Information and Knowledge Management},
pp. \bfpage{2621}--\blpage{2628}
(\byear{2020})
\end{bchapter}
\endbibitem

\bibitem{wang2020beyond}
\begin{bchapter}
\bauthor{\bsnm{Wang}, \binits{W.}},
\bauthor{\bsnm{Zhang}, \binits{W.}},
\bauthor{\bsnm{Liu}, \binits{S.}},
\bauthor{\bsnm{Liu}, \binits{Q.}},
\bauthor{\bsnm{Zhang}, \binits{B.}},
\bauthor{\bsnm{Lin}, \binits{L.}},
\bauthor{\bsnm{Zha}, \binits{H.}}:
\bctitle{Beyond clicks: Modeling multi-relational item graph for session-based target behavior prediction}.
In: \bbtitle{Proceedings of The Web Conference 2020},
pp. \bfpage{3056}--\blpage{3062}
(\byear{2020})
\end{bchapter}
\endbibitem

\bibitem{mcauley2015image}
\begin{bchapter}
\bauthor{\bsnm{McAuley}, \binits{J.}},
\bauthor{\bsnm{Targett}, \binits{C.}},
\bauthor{\bsnm{Shi}, \binits{Q.}},
\bauthor{\bsnm{Van Den~Hengel}, \binits{A.}}:
\bctitle{Image-based recommendations on styles and substitutes}.
In: \bbtitle{Proceedings of the 38th International ACM SIGIR Conference on Research and Development in Information Retrieval},
pp. \bfpage{43}--\blpage{52}
(\byear{2015})
\end{bchapter}
\endbibitem

\bibitem{mobasher2002using}
\begin{bchapter}
\bauthor{\bsnm{Mobasher}, \binits{B.}},
\bauthor{\bsnm{Dai}, \binits{H.}},
\bauthor{\bsnm{Luo}, \binits{T.}},
\bauthor{\bsnm{Nakagawa}, \binits{M.}}:
\bctitle{Using sequential and non-sequential patterns in predictive web usage mining tasks}.
In: \bbtitle{IEEE International Conference on Data Mining},
pp. \bfpage{669}--\blpage{672}
(\byear{2002}).
\bcomment{IEEE}
\end{bchapter}
\endbibitem

\bibitem{wu2013personalized}
\begin{bchapter}
\bauthor{\bsnm{Wu}, \binits{X.}},
\bauthor{\bsnm{Liu}, \binits{Q.}},
\bauthor{\bsnm{Chen}, \binits{E.}},
\bauthor{\bsnm{He}, \binits{L.}},
\bauthor{\bsnm{Lv}, \binits{J.}},
\bauthor{\bsnm{Cao}, \binits{C.}},
\bauthor{\bsnm{Hu}, \binits{G.}}:
\bctitle{Personalized next-song recommendation in online karaokes}.
In: \bbtitle{Proceedings of the 7th ACM Conference on Recommender Systems},
pp. \bfpage{137}--\blpage{140}
(\byear{2013})
\end{bchapter}
\endbibitem

\bibitem{le2016modeling}
\begin{bchapter}
\bauthor{\bsnm{Le}, \binits{D.-T.}},
\bauthor{\bsnm{Fang}, \binits{Y.}},
\bauthor{\bsnm{Lauw}, \binits{H.W.}}:
\bctitle{Modeling sequential preferences with dynamic user and context factors}.
In: \bbtitle{Proceedings of the Joint European Conference on Machine Learning and Knowledge Discovery in Databases},
pp. \bfpage{145}--\blpage{161}
(\byear{2016}).
\bcomment{Springer}
\end{bchapter}
\endbibitem

\bibitem{shani2005mdp}
\begin{botherref}
\oauthor{\bsnm{Shani}, \binits{G.}},
\oauthor{\bsnm{Heckerman}, \binits{D.}},
\oauthor{\bsnm{Brafman}, \binits{R.I.}},
\oauthor{\bsnm{Boutilier}, \binits{C.}}:
An mdp-based recommender system.
Journal of Machine Learning Research
\textbf{6}(9)
(2005)
\end{botherref}
\endbibitem

\bibitem{chen2012playlist}
\begin{bchapter}
\bauthor{\bsnm{Chen}, \binits{S.}},
\bauthor{\bsnm{Moore}, \binits{J.L.}},
\bauthor{\bsnm{Turnbull}, \binits{D.}},
\bauthor{\bsnm{Joachims}, \binits{T.}}:
\bctitle{Playlist prediction via metric embedding}.
In: \bbtitle{Proceedings of the 18th ACM SIGKDD International Conference on Knowledge Discovery and Data Mining},
pp. \bfpage{714}--\blpage{722}
(\byear{2012})
\end{bchapter}
\endbibitem

\bibitem{donkers2017sequential}
\begin{bchapter}
\bauthor{\bsnm{Donkers}, \binits{T.}},
\bauthor{\bsnm{Loepp}, \binits{B.}},
\bauthor{\bsnm{Ziegler}, \binits{J.}}:
\bctitle{Sequential user-based recurrent neural network recommendations}.
In: \bbtitle{Proceedings of the 11th ACM Conference on Recommender Systems},
pp. \bfpage{152}--\blpage{160}
(\byear{2017})
\end{bchapter}
\endbibitem

\bibitem{quadrana2017personalizing}
\begin{bchapter}
\bauthor{\bsnm{Quadrana}, \binits{M.}},
\bauthor{\bsnm{Karatzoglou}, \binits{A.}},
\bauthor{\bsnm{Hidasi}, \binits{B.}},
\bauthor{\bsnm{Cremonesi}, \binits{P.}}:
\bctitle{Personalizing session-based recommendations with hierarchical recurrent neural networks}.
In: \bbtitle{Proceedings of the 11th ACM Conference on Recommender Systems},
pp. \bfpage{130}--\blpage{137}
(\byear{2017})
\end{bchapter}
\endbibitem

\bibitem{hidasi2016parallel}
\begin{bchapter}
\bauthor{\bsnm{Hidasi}, \binits{B.}},
\bauthor{\bsnm{Quadrana}, \binits{M.}},
\bauthor{\bsnm{Karatzoglou}, \binits{A.}},
\bauthor{\bsnm{Tikk}, \binits{D.}}:
\bctitle{Parallel recurrent neural network architectures for feature-rich session-based recommendations}.
In: \bbtitle{Proceedings of the 10th ACM Conference on Recommender Systems},
pp. \bfpage{241}--\blpage{248}
(\byear{2016})
\end{bchapter}
\endbibitem

\bibitem{hidasi2015session}
\begin{bchapter}
\bauthor{\bsnm{Hidasi}, \binits{B.}},
\bauthor{\bsnm{Karatzoglou}, \binits{A.}},
\bauthor{\bsnm{Baltrunas}, \binits{L.}},
\bauthor{\bsnm{Tikk}, \binits{D.}}:
\bctitle{Session-based recommendations with recurrent neural networks}.
In: \bbtitle{Proceedings of the 4th International Conference on Learning Representations}
(\byear{2016})
\end{bchapter}
\endbibitem

\bibitem{li2017neural}
\begin{bchapter}
\bauthor{\bsnm{Li}, \binits{J.}},
\bauthor{\bsnm{Ren}, \binits{P.}},
\bauthor{\bsnm{Chen}, \binits{Z.}},
\bauthor{\bsnm{Ren}, \binits{Z.}},
\bauthor{\bsnm{Lian}, \binits{T.}},
\bauthor{\bsnm{Ma}, \binits{J.}}:
\bctitle{Neural attentive session-based recommendation}.
In: \bbtitle{Proceedings of the 26th ACM International Conference on Conference on Information and Knowledge Management},
pp. \bfpage{1419}--\blpage{1428}
(\byear{2017})
\end{bchapter}
\endbibitem

\bibitem{liu2018stamp}
\begin{bchapter}
\bauthor{\bsnm{Liu}, \binits{Q.}},
\bauthor{\bsnm{Zeng}, \binits{Y.}},
\bauthor{\bsnm{Mokhosi}, \binits{R.}},
\bauthor{\bsnm{Zhang}, \binits{H.}}:
\bctitle{Stamp: short-term attention/memory priority model for session-based recommendation}.
In: \bbtitle{Proceedings of the 24th ACM SIGKDD International Conference on Knowledge Discovery and Data Mining},
pp. \bfpage{1831}--\blpage{1839}
(\byear{2018})
\end{bchapter}
\endbibitem

\bibitem{qiu2019rethinking}
\begin{bchapter}
\bauthor{\bsnm{Qiu}, \binits{R.}},
\bauthor{\bsnm{Li}, \binits{J.}},
\bauthor{\bsnm{Huang}, \binits{Z.}},
\bauthor{\bsnm{Yin}, \binits{H.}}:
\bctitle{Rethinking the item order in session-based recommendation with graph neural networks}.
In: \bbtitle{Proceedings of the 28th ACM International Conference on Information and Knowledge Management},
pp. \bfpage{579}--\blpage{588}
(\byear{2019})
\end{bchapter}
\endbibitem

\bibitem{wu2019session}
\begin{bchapter}
\bauthor{\bsnm{Wu}, \binits{S.}},
\bauthor{\bsnm{Tang}, \binits{Y.}},
\bauthor{\bsnm{Zhu}, \binits{Y.}},
\bauthor{\bsnm{Wang}, \binits{L.}},
\bauthor{\bsnm{Xie}, \binits{X.}},
\bauthor{\bsnm{Tan}, \binits{T.}}:
\bctitle{Session-based recommendation with graph neural networks}.
In: \bbtitle{Proceedings of the 33rd AAAI Conference on Artificial Intelligence},
vol. \bseriesno{33},
pp. \bfpage{346}--\blpage{353}
(\byear{2019})
\end{bchapter}
\endbibitem

\bibitem{li2015gated}
\begin{bchapter}
\bauthor{\bsnm{Li}, \binits{Y.}},
\bauthor{\bsnm{Tarlow}, \binits{D.}},
\bauthor{\bsnm{Brockschmidt}, \binits{M.}},
\bauthor{\bsnm{Zemel}, \binits{R.}}:
\bctitle{Gated graph sequence neural networks}.
In: \bbtitle{Proceedings of the 3rd International Conference on Learning Representations}
(\byear{2015})
\end{bchapter}
\endbibitem

\bibitem{xu2019graph}
\begin{bchapter}
\bauthor{\bsnm{Xu}, \binits{C.}},
\bauthor{\bsnm{Zhao}, \binits{P.}},
\bauthor{\bsnm{Liu}, \binits{Y.}},
\bauthor{\bsnm{Sheng}, \binits{V.S.}},
\bauthor{\bsnm{Xu}, \binits{J.}},
\bauthor{\bsnm{Zhuang}, \binits{F.}},
\bauthor{\bsnm{Fang}, \binits{J.}},
\bauthor{\bsnm{Zhou}, \binits{X.}}:
\bctitle{Graph contextualized self-attention network for session-based recommendation}.
In: \bbtitle{Proceedings of the 28th International Joint Conference on Artificial Intelligence},
vol. \bseriesno{19},
pp. \bfpage{3940}--\blpage{3946}
(\byear{2019})
\end{bchapter}
\endbibitem

\bibitem{chen2020handling}
\begin{bchapter}
\bauthor{\bsnm{Chen}, \binits{T.}},
\bauthor{\bsnm{Wong}, \binits{R.C.-W.}}:
\bctitle{Handling information loss of graph neural networks for session-based recommendation}.
In: \bbtitle{Proceedings of the 26th ACM SIGKDD International Conference on Knowledge Discovery and Data Mining},
pp. \bfpage{1172}--\blpage{1180}
(\byear{2020})
\end{bchapter}
\endbibitem

\bibitem{xia2021self}
\begin{bchapter}
\bauthor{\bsnm{Xia}, \binits{X.}},
\bauthor{\bsnm{Yin}, \binits{H.}},
\bauthor{\bsnm{Yu}, \binits{J.}},
\bauthor{\bsnm{Wang}, \binits{Q.}},
\bauthor{\bsnm{Cui}, \binits{L.}},
\bauthor{\bsnm{Zhang}, \binits{X.}}:
\bctitle{Self-supervised hypergraph convolutional networks for session-based recommendation}.
In: \bbtitle{Proceedings of the 35th AAAI Conference on Artificial Intelligence},
vol. \bseriesno{35},
pp. \bfpage{4503}--\blpage{4511}
(\byear{2021})
\end{bchapter}
\endbibitem

\bibitem{wang2020global}
\begin{bchapter}
\bauthor{\bsnm{Wang}, \binits{Z.}},
\bauthor{\bsnm{Wei}, \binits{W.}},
\bauthor{\bsnm{Cong}, \binits{G.}},
\bauthor{\bsnm{Li}, \binits{X.-L.}},
\bauthor{\bsnm{Mao}, \binits{X.-L.}},
\bauthor{\bsnm{Qiu}, \binits{M.}}:
\bctitle{Global context enhanced graph neural networks for session-based recommendation}.
In: \bbtitle{Proceedings of the 43rd International ACM SIGIR Conference on Research and Development in Information Retrieval},
pp. \bfpage{169}--\blpage{178}
(\byear{2020})
\end{bchapter}
\endbibitem

\bibitem{xia2021self1}
\begin{bchapter}
\bauthor{\bsnm{Xia}, \binits{X.}},
\bauthor{\bsnm{Yin}, \binits{H.}},
\bauthor{\bsnm{Yu}, \binits{J.}},
\bauthor{\bsnm{Shao}, \binits{Y.}},
\bauthor{\bsnm{Cui}, \binits{L.}}:
\bctitle{Self-supervised graph co-training for session-based recommendation}.
In: \bbtitle{Proceedings of the 30th ACM International Conference on Information and Knowledge Management},
pp. \bfpage{2180}--\blpage{2190}
(\byear{2021})
\end{bchapter}
\endbibitem

\bibitem{xia2022multi}
\begin{botherref}
\oauthor{\bsnm{Xia}, \binits{L.}},
\oauthor{\bsnm{Huang}, \binits{C.}},
\oauthor{\bsnm{Xu}, \binits{Y.}},
\oauthor{\bsnm{Pei}, \binits{J.}}:
Multi-behavior sequential recommendation with temporal graph transformer.
IEEE Transactions on Knowledge and Data Engineering
(2022)
\end{botherref}
\endbibitem

\bibitem{xia2021graph}
\begin{bchapter}
\bauthor{\bsnm{Xia}, \binits{L.}},
\bauthor{\bsnm{Xu}, \binits{Y.}},
\bauthor{\bsnm{Huang}, \binits{C.}},
\bauthor{\bsnm{Dai}, \binits{P.}},
\bauthor{\bsnm{Bo}, \binits{L.}}:
\bctitle{Graph meta network for multi-behavior recommendation}.
In: \bbtitle{Proceedings of the 44th International ACM SIGIR Conference on Research and Development in Information Retrieval},
pp. \bfpage{757}--\blpage{766}
(\byear{2021})
\end{bchapter}
\endbibitem

\bibitem{xia2021knowledge}
\begin{bchapter}
\bauthor{\bsnm{Xia}, \binits{L.}},
\bauthor{\bsnm{Huang}, \binits{C.}},
\bauthor{\bsnm{Xu}, \binits{Y.}},
\bauthor{\bsnm{Dai}, \binits{P.}},
\bauthor{\bsnm{Zhang}, \binits{X.}},
\bauthor{\bsnm{Yang}, \binits{H.}},
\bauthor{\bsnm{Pei}, \binits{J.}},
\bauthor{\bsnm{Bo}, \binits{L.}}:
\bctitle{Knowledge-enhanced hierarchical graph transformer network for multi-behavior recommendation}.
In: \bbtitle{Proceedings of the 35th AAAI Conference on Artificial Intelligence},
vol. \bseriesno{35},
pp. \bfpage{4486}--\blpage{4493}
(\byear{2021})
\end{bchapter}
\endbibitem

\bibitem{meng2020incorporating}
\begin{bchapter}
\bauthor{\bsnm{Meng}, \binits{W.}},
\bauthor{\bsnm{Yang}, \binits{D.}},
\bauthor{\bsnm{Xiao}, \binits{Y.}}:
\bctitle{Incorporating user micro-behaviors and item knowledge into multi-task learning for session-based recommendation}.
In: \bbtitle{Proceedings of the 43rd International ACM SIGIR Conference on Research and Development in Information Retrieval},
pp. \bfpage{1091}--\blpage{1100}
(\byear{2020})
\end{bchapter}
\endbibitem

\bibitem{wan2018representing}
\begin{bchapter}
\bauthor{\bsnm{Wan}, \binits{M.}},
\bauthor{\bsnm{Wang}, \binits{D.}},
\bauthor{\bsnm{Liu}, \binits{J.}},
\bauthor{\bsnm{Bennett}, \binits{P.}},
\bauthor{\bsnm{McAuley}, \binits{J.}}:
\bctitle{Representing and recommending shopping baskets with complementarity, compatibility and loyalty}.
In: \bbtitle{Proceedings of the 27th ACM International Conference on Information and Knowledge Management},
pp. \bfpage{1133}--\blpage{1142}
(\byear{2018})
\end{bchapter}
\endbibitem

\bibitem{zhao2017improving}
\begin{bchapter}
\bauthor{\bsnm{Zhao}, \binits{T.}},
\bauthor{\bsnm{McAuley}, \binits{J.}},
\bauthor{\bsnm{Li}, \binits{M.}},
\bauthor{\bsnm{King}, \binits{I.}}:
\bctitle{Improving recommendation accuracy using networks of substitutable and complementary products}.
In: \bbtitle{International Joint Conference on Neural Networks},
pp. \bfpage{3649}--\blpage{3655}
(\byear{2017}).
\bcomment{IEEE}
\end{bchapter}
\endbibitem

\bibitem{chen2020try}
\begin{bchapter}
\bauthor{\bsnm{Chen}, \binits{T.}},
\bauthor{\bsnm{Yin}, \binits{H.}},
\bauthor{\bsnm{Ye}, \binits{G.}},
\bauthor{\bsnm{Huang}, \binits{Z.}},
\bauthor{\bsnm{Wang}, \binits{Y.}},
\bauthor{\bsnm{Wang}, \binits{M.}}:
\bctitle{Try this instead: Personalized and interpretable substitute recommendation}.
In: \bbtitle{Proceedings of the 43rd International ACM SIGIR Conference on Research and Development in Information Retrieval},
pp. \bfpage{891}--\blpage{900}
(\byear{2020})
\end{bchapter}
\endbibitem

\bibitem{zhang2021learning}
\begin{barticle}
\bauthor{\bsnm{Zhang}, \binits{W.}},
\bauthor{\bsnm{Chen}, \binits{Z.}},
\bauthor{\bsnm{Zha}, \binits{H.}},
\bauthor{\bsnm{Wang}, \binits{J.}}:
\batitle{Learning from substitutable and complementary relations for graph-based sequential product recommendation}.
\bjtitle{ACM Transactions on Information Systems (TOIS)}
\bvolume{40}(\bissue{2}),
\bfpage{1}--\blpage{28}
(\byear{2021})
\end{barticle}
\endbibitem

\bibitem{zheng2020robust}
\begin{bchapter}
\bauthor{\bsnm{Zheng}, \binits{C.}},
\bauthor{\bsnm{Zong}, \binits{B.}},
\bauthor{\bsnm{Cheng}, \binits{W.}},
\bauthor{\bsnm{Song}, \binits{D.}},
\bauthor{\bsnm{Ni}, \binits{J.}},
\bauthor{\bsnm{Yu}, \binits{W.}},
\bauthor{\bsnm{Chen}, \binits{H.}},
\bauthor{\bsnm{Wang}, \binits{W.}}:
\bctitle{Robust graph representation learning via neural sparsification}.
In: \bbtitle{Proceedings of the 12th International Conference on Machine Learning},
pp. \bfpage{11458}--\blpage{11468}
(\byear{2020}).
\bcomment{PMLR}
\end{bchapter}
\endbibitem

\bibitem{luo2021learning}
\begin{bchapter}
\bauthor{\bsnm{Luo}, \binits{D.}},
\bauthor{\bsnm{Cheng}, \binits{W.}},
\bauthor{\bsnm{Yu}, \binits{W.}},
\bauthor{\bsnm{Zong}, \binits{B.}},
\bauthor{\bsnm{Ni}, \binits{J.}},
\bauthor{\bsnm{Chen}, \binits{H.}},
\bauthor{\bsnm{Zhang}, \binits{X.}}:
\bctitle{Learning to drop: Robust graph neural network via topological denoising}.
In: \bbtitle{Proceedings of the 14th ACM International Conference on Web Search and Data Mining},
pp. \bfpage{779}--\blpage{787}
(\byear{2021})
\end{bchapter}
\endbibitem

\bibitem{zhang2019inferring}
\begin{bchapter}
\bauthor{\bsnm{Zhang}, \binits{S.}},
\bauthor{\bsnm{Yin}, \binits{H.}},
\bauthor{\bsnm{Wang}, \binits{Q.}},
\bauthor{\bsnm{Chen}, \binits{T.}},
\bauthor{\bsnm{Chen}, \binits{H.}},
\bauthor{\bsnm{Nguyen}, \binits{Q.V.H.}}:
\bctitle{Inferring substitutable products with deep network embedding}.
In: \bbtitle{Proceedings of the 28th International Joint Conference on Artificial Intelligence},
pp. \bfpage{4306}--\blpage{4312}
(\byear{2019})
\end{bchapter}
\endbibitem

\bibitem{sarwar2001item}
\begin{bchapter}
\bauthor{\bsnm{Sarwar}, \binits{B.}},
\bauthor{\bsnm{Karypis}, \binits{G.}},
\bauthor{\bsnm{Konstan}, \binits{J.}},
\bauthor{\bsnm{Riedl}, \binits{J.}}:
\bctitle{Item-based collaborative filtering recommendation algorithms}.
In: \bbtitle{Proceedings of the 10th International Conference on World Wide Web},
pp. \bfpage{285}--\blpage{295}
(\byear{2001})
\end{bchapter}
\endbibitem

\bibitem{ren2019repeatnet}
\begin{bchapter}
\bauthor{\bsnm{Ren}, \binits{P.}},
\bauthor{\bsnm{Chen}, \binits{Z.}},
\bauthor{\bsnm{Li}, \binits{J.}},
\bauthor{\bsnm{Ren}, \binits{Z.}},
\bauthor{\bsnm{Ma}, \binits{J.}},
\bauthor{\bsnm{De~Rijke}, \binits{M.}}:
\bctitle{Repeatnet: A repeat aware neural recommendation machine for session-based recommendation}.
In: \bbtitle{Proceedings of the 33rd AAAI Conference on Artificial Intelligence},
vol. \bseriesno{33},
pp. \bfpage{4806}--\blpage{4813}
(\byear{2019})
\end{bchapter}
\endbibitem

\end{thebibliography}

\end{document}